\documentclass[twocolumn,floatfix]{aastex63}

\usepackage{color}

\usepackage{float}
\usepackage{multirow}
\usepackage{appendix}
\usepackage{amsmath}

\newcommand\kms{km\,s$^{-1}$}

\received{20/03/2021}
\accepted{31/05/2021}%\today}
\submitjournal{ApJ}

\shorttitle{Probing jets from young embedded sources}
\shortauthors{Erkal et al.}
\graphicspath{{./}{figures/}}
\begin{document}

\title{Probing jets from young embedded sources: clues from \textit{HST} near-IR [Fe II] images}

\correspondingauthor{Jessica Erkal}
\email{jessica.erkal@ucdconnect.ie}

\author[0000-0002-8476-1389]{Jessica Erkal}
\affiliation{School of Physics, University College Dublin, Belfield, Dublin 4, Ireland}

\author[0000-0002-9190-0113]{Brunella Nisini}
\affiliation{INAF, Osservatorio Astronomico di Roma, via Frascati 33,  00078,  Monte Porzio Catone,  Italy}

\author[0000-0002-2210-202X]{Deirdre Coffey}
\affiliation{School of Physics, University College Dublin, Belfield, Dublin 4, Ireland}

\author[0000-0001-5776-9476]{Francesca Bacciotti}
\affiliation{INAF, Osservatorio Astrofisico di Arcetri, Largo E. Fermi 5, 50125, Firenze, Italy}

\author[0000-0002-5380-549X]{Patrick Hartigan}
\affiliation{Physics and Astronomy Dept., Rice University, 6100 S. Main, Houston, TX 77005-1892}

\author[0000-0002-0666-3847]{Simone Antoniucci}
\affiliation{INAF, Osservatorio Astronomico di Roma, via Frascati 33,  00078,  Monte Porzio Catone,  Italy}

\author[0000-0002-7035-8513]{Teresa Giannini}
\affiliation{INAF, Osservatorio Astronomico di Roma, via Frascati 33,  00078,  Monte Porzio Catone,  Italy}

\author[0000-0001-6496-0252]{Jochen Eisl\"offel}
\affiliation{Th\"uringer Landessternwarte, Sternwarte 5, 07778, Tautenburg, Germany}

\author[0000-0003-3562-262X]{Carlo Felice Manara}
\affiliation{European Southern Observatory, Karl-Schwarzschild-Strasse 2, 85748, Garching bei M\"unchen, Germany}
%%%%%%%%%%%%%%%%%%%%%%%%%%%%%%%%%%%%%%%%%%%%%%%%%%%%%%%%%%%%%%%%%%%%%%%%%%%%%%%%
\begin{abstract}

We present near-infrared [Fe II] images of four Class 0/I jets (HH 1/2, HH 34, HH 111, HH 46/47) observed with the Hubble Space Telescope Wide Field Camera 3. The unprecedented angular resolution allows us to measure proper motions, jet widths and trajectories, and extinction along the jets. In all cases, we detect the counter-jet which was barely visible or invisible at shorter wavelengths. We measure tangential velocities of a few hundred km/s, consistent with previous HST measurements over 10 years ago. We measure the jet width as close as a few tens of au from the star, revealing high collimations of about 2 degrees for HH 1, HH 34, HH 111 and about 8 degrees for HH 46, all of which are preserved up to large distances. For HH 34, we find evidence of a larger initial opening angle of about 7 degrees. Measurement of knot positions reveals deviations in trajectory of both the jet and counter-jet of all sources. Analysis of asymmetries in the inner knot positions for HH 111 suggests the presence of a low mass stellar companion at separation 20-30 au. Finally, we find extinction values of 15-20 mag near the source which gradually decreases moving downstream along the jet. These observations have allowed us to study the counter-jet at unprecedented high angular resolution, and will be a valuable reference for planning future JWST mid-infrared observations which will peer even closer into the jet engine.

\end{abstract}

%%%%%%%%%%%%%%%%%%%%%%%%%%%%%%%%%%%%%%%%%%%%%%%%%%%%%%%%%%%%%%%%%%%%%%%%%%%%%%%%

\section{Introduction} 
\label{sec:intro}

%\begin{itemize}
%    \item Understanding of jets mostly on direct observations of more evolved T~Tauri stars
%    \item Need to observe younger stars to understand original formation and how properties evolve with time
%    \item Extinction using iron lines and measure other properties eg collimation, mass flux and maybe density
%    \item Important archival reference for future observations incl JWST - legacy value
%\end{itemize}

Bipolar jets are a key ingredient in the star formation process, but the role they play is not yet fully understood \citep{Frank2014}. They are found to transport significant amounts of mass and momentum away from the newly forming star, but theoretical models still struggle to agree on their basic formation mechanism \citep [e.g.][]{Ferreira2006}.

Most research to-date has focused on more evolved (Class II) sources which have already cleared their surrounding envelopes and therefore are easier to observe with optical instrumentation \citep{Frank2014}. However, to understand the earlier evolutionary stages, we must also study jets from less evolved (Class 0/I) sources. These sources are still highly embedded in their dusty envelope and dense cloud of material and the resulting large extinction prevents an accurate determination of their physical properties close to the star. Consequently, detailed studies of extended jets from these objects are usually performed only far from the central source (i.e. $\ga$ 10$"$), where the problem of extinction becomes less severe. In these regions, however, the jet has already interacted with the ambient medium through multiple shocks, loosing the pristine information about its acceleration mechanism.

Hence, studies of the jet base require observations at longer wavelengths, which can penetrate the circumstellar envelope, combined with an accurate estimate of the associated extinction to allow correct interpretation of plasma conditions. In this respect, [\ion{Fe}{2}] lines in the near-IR have been shown to be important tracers of embedded jets \citep[e.g.][]{Davis2003,Nisini2002}. Ground-based observations of the bright 1.64 and 1.25$\mu$m transitions, in particular, have been widely used to probe Class 0/I dense jets and  to measure their extinction \citep[e.g.][]{Nisini2005,Davis2011,GarciaLopez2010}. 

Sub-arcsecond spatial resolution observations are needed to allow identification of the morphology of internal shock fronts (a.k.a. knots) within the jet stream, and to resolve their widths, allowing jet collimation to be measured. However, ground-based IR facilities can provide only limited spatial resolution on embedded targets, because such sources do not constitute suitable natural guide stars for adaptive optics. Observations from space are therefore needed to reach the required resolution.

Here we present Hubble Space Telescope (\textit{HST}) Wide Field Camera 3 (WFC3) images in [\ion{Fe}{2}] lines of four well known Class 0/I sources, namely HH~1/2, HH~34, HH~46/47 and HH~111, that have been extensively studied by {\it HST} in the optical through multi-epoch imaging in the H$\alpha$ and [\ion{S}{2}] emission. 
In particular, comprehensive studies of HH~1 \citep{Reipurth2000a, Bally2002}, HH~34 and HH~47 \citep{Reipurth2002,Hartigan2005,Hartigan2011}, and HH~111 \citep{Hartigan2001, NoreigaCrespo2011} have addressed their morphological changes, brightness variations and proper motions, mainly in the outer, optical bright part of these jets. Previous {\it HST} Near Infrared Camera and Multi-Object Spectrometer (NICMOS) images of these sources \citep{Reipurth2000a, Reipurth2000b} have revealed details hidden in optical observations, tracing the jet emission closer to the driving source than previously possible. However, these previous NIR images, obtained with broad-band filters, were not suited to infer properties of the inner jets, where the emission from source continuum nebulosity dominates over the jet line emission.   

In our study, we have acquired narrow band images centred on [\ion{Fe}{2}] 1.64 and 1.25~$\mu$m together with images taken in continuum emission at adjacent wavelengths, in order to perform an optimum continuum subtraction and analyse the jet emission as close as possible to the source. These images allow us to study jet physics by examining the initial jet collimation and velocities, extinction along the jets, and counter-jet asymmetries. These observations will also have a legacy value for future mid-IR observations on these outflow with the James Webb Space Telescope (JWST), which opens an important wavelength window on jet launching in Class~0/I sources. 

The paper is structured in the following way: Section 2 presents the image acquisition and data reduction performed in order to obtain flux calibrated and continuum subtracted images of the four outflows. Section 3 presents the general morphology of the [\ion{Fe}{2}] emission, in comparison with previous optical images of the same jets, and describes the analysis performed on the images. In Section 4 we discuss our results and we give our main conclusions in Section 5. 

\begin{table*}[ht]
%\centering
\caption{Properties of the four targets}
\begin{tabular}{lcccccccc}
\hline
Target & Class & RA & DEC & Ref. & Distance & Jet PA & Inclination & Ref. \\
 &  &  & &  & (pc) & ($^{\circ}$) & ($^{\circ}$) & \\ \hline
HH~1/2 & 0 & 05:36:22.840 & -06:46:06.20 & 1 & 383 & 325.5 & 10 & 5,6 \\
HH~34 & I & 05:35:29.846 & -06:26:58.08 & 2 & 383 & 165.7 & 34 & 7,6 \\
HH~46/47 & I & 08:25:43.800 & -51:00:36.00 & 3 & 450 & 52.3 & 34 & 8\\
HH~111 & I & 05:51:46.254 & +02:48:29.65 & 4 & 400 & 277.3 & 10 & 4,9 \\ \hline
\label{table:targets}
\end{tabular}

%$^a$ References for the coordinates of the driving source: $^b$ References for the inclination (given with respect to the plane of the sky) and distance: 
(1)\citet{Rodriguez2000}; (2) \citet{Rodriguez2014}; (3) \citet{Arce2013}; (4) \citet{Lee2016}; (5) \citet{Bally2002}; (6) \citet{Grossschedl2018}; (7) \citet{Raga2012}; (8) \citet{Reipurth2000b}; (9) \citet{Raga2002}
\end{table*}
%add ra dec of targets

%In section 2 we describe the observations and data reduction; in section 3 we describe the results on jet morphology at large scales and in the inner jet regions, proper motions and extinction values along the jets; in section 4 we discuss models of the HH~111 jet wiggling and ... \jess{tbd}

%will find the rest of these values tomorrow, need to add references

%%%%%%%%%%%%%%%%%%%%%%%%%%%%%%%%%%%%%%%%%%%%%%%%%%%%%%%%%%%%%%%%%%%%%%%%%%%%%%%%
\section{Observations \& Data Reduction} \label{sec:observations}

%\begin{itemize}
  %  \item Four bright jets from Class~0/I sources
    %\item HST WFC3 in optical and IR narrow band filters
    %item High angular resolution to try to observe close to the source

   % \item Astrometry - aligning and shifting
    %\item Photometry - check flux calibration 
%\end{itemize}

\subsection{Observations}

%Using the Hubble Space Telescope (\textit{HST}) Wide Field Camera 3 (WFC3) 
Using \textit{HST} WFC3 we observe four Class~0/I jets (HH~1/2, HH~34 (d=383 pc), HH~111 (d=400 pc) in the Orion Nebula and HH~46/47 in the Gum nebula, (d=450 pc)(Program ID: 15178, PI: B. Nisini). The WFC3 field of view (FoV) is 136$\arcsec$ $\times$ 123$\arcsec$ in the IR and 162$\arcsec$ $\times$ 162$\arcsec$ in the UVIS channel, while the spatial sampling is 0$\farcs$13 and 0$\farcs$04, respectively. For three sources (HH~34, HH~46/47 and HH~111), a single FoV position was sufficient to include the relevant sections of the jet, while for the fourth target (HH~1/2) two overlapping FoV positions were required to achieve full spatial coverage that includes both the HH1 and HH2 bow shocks. In all cases, the jet was positioned diagonally across the field of view to maximise spatial coverage. Images were obtained in three narrowband filters: F126N and F164N (IR channel); and F631N (UVIS channel). In addition, images in adjacent narrowband filters (F130N, F167N and F645N) were obtained to facilitate continuum subtraction. Exposure times vary between $\approx$ 500-2000 seconds in each filter. Details of the observations are summarised in Table  \ref{table:observations}. 

\begin{table*}[ht]
\centering
\caption{Summary of \textit{HST} WFC3 observation dates and exposure times for each filter used. We also include the central wavelength and FWHM of each filter.}
\begin{tabular}{cccccccc}
\hline
Target & Date & F126N & F130N & F164N & F167N & F631N & F645N \\
 &  & (s) & (s) & (s) & (s) & (s) & (s) \\ \hline
HH~1 (North) & 20 Jan 2019 & 1105 & 455 & 905 & 705 & 700 & 746 \\
HH~1 (South) & 20 Jan 2019 & 1105 & 555 & 805 & 705 & 700 & 700 \\
HH~34 & 31 Jan 2019 & 1958 & 1305 & 1958 & 1958 & 1764 & 1168 \\
HH~46 & 29 Mar 2019 & 1958 & 1605 & 1958 & 1958 & 1920 & 1385 \\
HH~111 & 28 Mar 2019 & 2108 & 1005 & 2108 & 1958 & 1761 & 1167 \\ \hline
\hline
$\lambda _{cen}$ (nm) & & 1258.5 & 1300.9 & 1645.1 & 1667.1 & 630.3 & 645.3 \\
FWHM (\AA) & & 151.19 & 156.28 & 208.5 & 210.85 & 61.33 & 85.12 \\ \hline
\label{table:observations}
\end{tabular}
\end{table*}

%\bru{need to add that for HH~46,111 and 34 a single esposure was sufficient to include the relevant sections of the outflows while two overlapping exposures have been acquired for HH~1, to accomodate both the jet, the HH~1 and the HH2 }
%\jess{Make a table with source, RA, DEC, observation date, filter and exposure time?}
%\bru{OK but filter list redundant since it is the same for all sources. We can have another small table with the filters characteristics (i.e. central wave, width, possible other contaminating lines)}

%Proposal ID 15178 Cycle 25 

\subsection{Data Reduction}
%Astrometry – align images with ccxymatch and ccmap OR by sub-pixel shifts (found using 2D gaussian fits on stars in FOV in QFitsView)

%For HH1 only – combining frames 1 \& 2. imcombine using offsets=wcs. Do combine before steps 4,5 and after steps 4,5 to compare
%\bru{I moved the above paragraph at this point because it is connected with the construction of the images}

%Check flux calibration for continuum sources - circular region around the star and getting the sum of the counts in the circle. star counts - sky counts and convert to erg/cm2/s/A using header keyword PHOTFLAM, convert to mag. Magnitudes were found to be within 0.5 mag of the 2MASS value

The data were calibrated through the standard \textit{HST} data reduction pipeline. IRAF software was used for further data reduction as follows. 

First, the continuum images were aligned to the line+continuum images. Using the \textit{ccxymatch} and \textit{ccmap} routines, pixel coordinates and RA/DEC coordinates for between 3-6 background stars in each image were provided to identify the required transformation to align the images. The new positions of the same stars in each image were compared using 2D Gaussian fits to ensure the images were indeed correctly aligned. Additionally, for HH~1 which was observed with two FoV positions, the frames were combined using the \textit{imcombine} routine with the \textit{offsets} keyword set to use the WCS header information to align each frame.

To flux calibrate the data, each image was multiplied by the \textit{PHOTFLAM} header keyword which converts the data units to flux units of ergs cm$^{-2}$ s$^{-1}$ \AA $^{-1}$. This calibration was checked to be accurate for continuum sources by converting the binned counts to magnitudes, for a selection of stars in the image. The magnitudes were found to be within 0.5 mag of the 2MASS catalogue value. The images were then multiplied by the filter width to obtain flux units of ergs/s/cm$^{2}$ in the line images.
This procedure gives correct results only if the filter transmission is uniform, so that the average filter throughput, used in the calibration with the PHOTFLAM parameter, does not significantly differ from the throughput at the exact wavelength where the lines were emitting. In the case of the adopted filters, the transmission curve is quite flat and we found that the average throughput value differs from the throughput at the line wavelength by only about 1\% . 

%\jess{The \textit{PHOTFLAM} parameter varies for each filter and the throughput within the filters can vary by up to 20\%, therefore it is crucial to check that using the \textit{PHOTFLAM} parameter to flux calibrate the images is sufficient. This was done by calculating the ratio of the \textit{PHOTFLAM} parameters for the 1.64~$\mu$m and 1.25~$\mu$m filters multiplied by the appropriate filter width, and comparing this to the ratio of the throughputs at the line wavelength multiplied by the inverse of the wavelengths. These two ratios were found to be roughly equal to about 1\% so we assume the \textit{PHOTFLAM} flux calibration to be correct.}

The aligned, flux-calibrated images were then continuum subtracted to remove the strong nebulosity around the source and observe the jet base close to the star. 
A direct subtraction of the continuum image, however, still leaves some residual nebulosity around the source, likely due to scattered line emission in the envelope cavities. 

%In some places, particularly close to the source, over-subtraction of the continuum was observed (i.e. dark negative features). Some residual nebulosity was also present. Faint emission lines and scattering in the continuum images are additional potential sources of noise in the continuum-subtracted images. 

%\bru{I think that in this section you have to stop here. The following, i.e. continuum subtraction and image ratios, should go into the subsequent analysis sections}

%%%%%%%%%%%%%%%%%%%%%%%%%%%%%%%%%%%%%%%%%%%%%%%%%%%%%%%%%%%%%%%%%%%%%%%%%%%%%%%%

\section{Results} \label{sec:results}

We present the large scale structure first, and then the inner jet channel closest to the star to examine the small scale structure. 

\subsection{Large Scale Structure}

\subsubsection{Jet morphology}

For each target, we present in Figures 1-4 the jet morphology using images in the [\ion{Fe}{2}] 1.64~$\mu$m emission line, as this line gives the richest detail. 
Images in [\ion{Fe}{2}] 1.25~$\mu$m and [\ion{O}{1}] $\lambda$6300 are shown in Appendix~\ref{appendix_a}. The [\ion{O}{1}] $\lambda$6300 images are very noisy, especially close to the source where the line emission is weak due to high extinction, and so will not be discussed further here. \textbf{In Appendix~\ref{appendix_b} we show the [\ion{O}{1}]/[\ion{Fe}{2}] images for regions where [\ion{O}{1}] emission is detected above the threshold (greater than 2 $\times$ RMS).} 

\begin{figure*}[ht]
    \centering
    \includegraphics[width=0.8\textwidth,keepaspectratio]{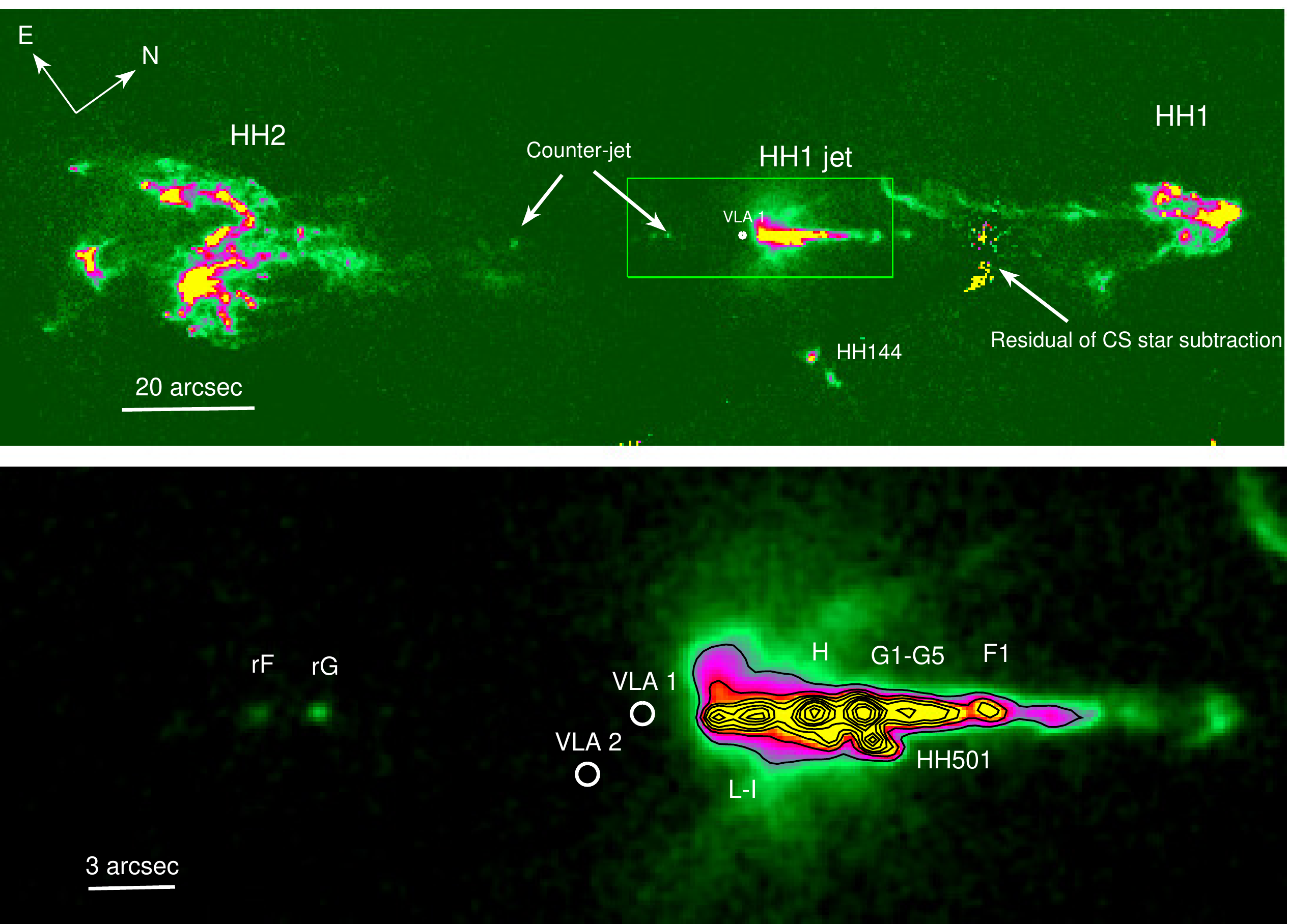}
    \caption{[\ion{Fe}{2}] 1.64$\mu$m continuum subtracted image of the HH~1/2 region. The bottom panel is an enlarged view of the inner HH~1 jet region, marked by a green box in the top panel. Contour levels are at 1, 2, 3, 4, 6, 8, 10, 15 and 20$\times$10$^{-19}$ \textbf{erg\,s$^{-1}$\,cm$^{-2}$\,\AA$^{-1}$\,pixel$^{-1}$.}}
    \label{fig:inner_regions_hh1}
\end{figure*}

\begin{figure*}[ht]
    \centering
    \includegraphics[width=0.8\textwidth,keepaspectratio]{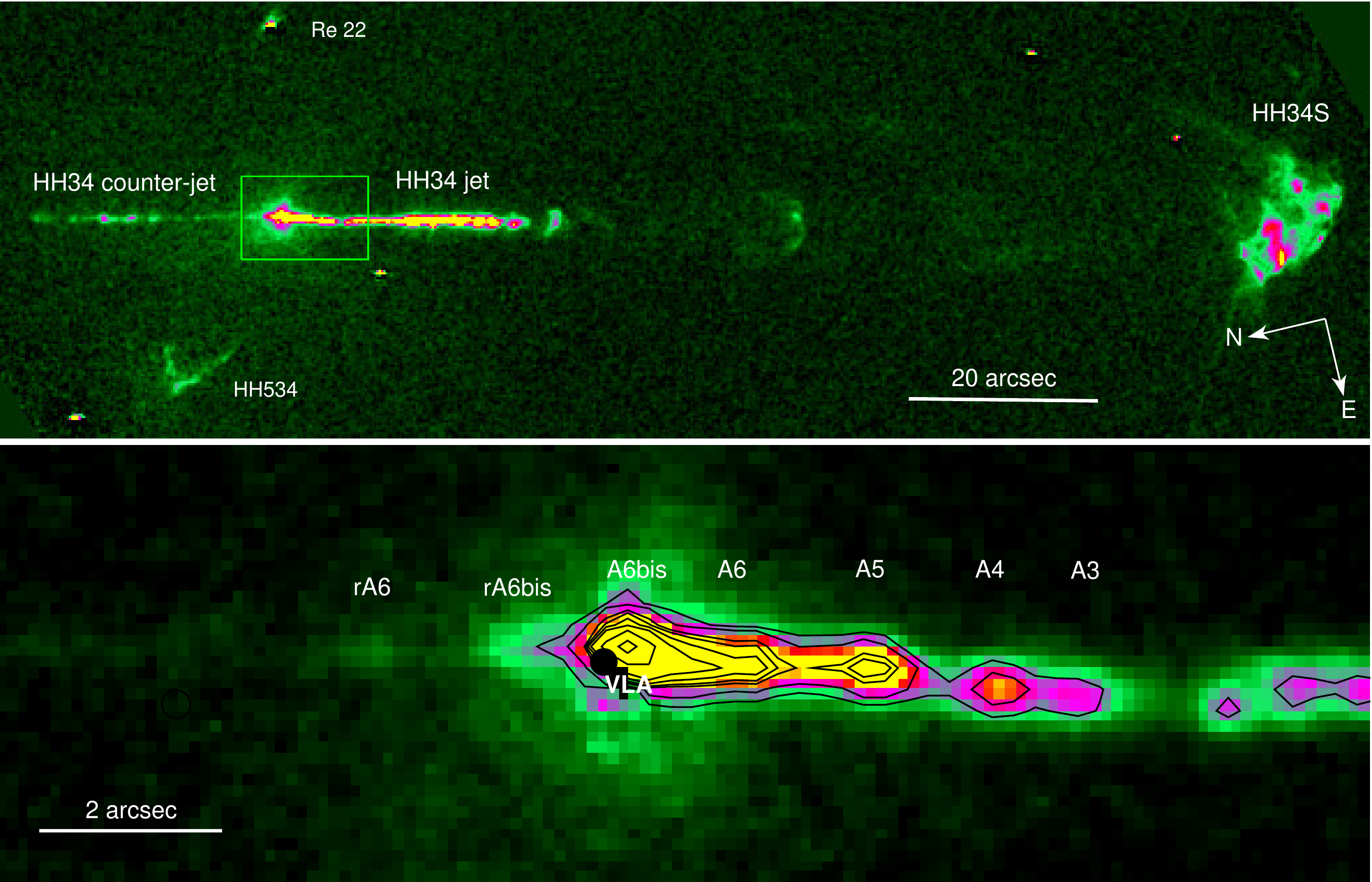}
    \caption{The same as Fig. \ref{fig:inner_regions_hh1} for the HH~34 region. Contour levels are at 1.2, 2, 4, 5, 7, 10, 20 and 40$\times$10$^{-19}$ \textbf{erg\,s$^{-1}$\,cm$^{-2}$\,\AA$^{-1}$\,pixel$^{-1}$.} }
    \label{fig:inner_regions_hh34}
\end{figure*}

\begin{figure*}[ht]
    \centering
    \includegraphics[width=0.8\textwidth,keepaspectratio]{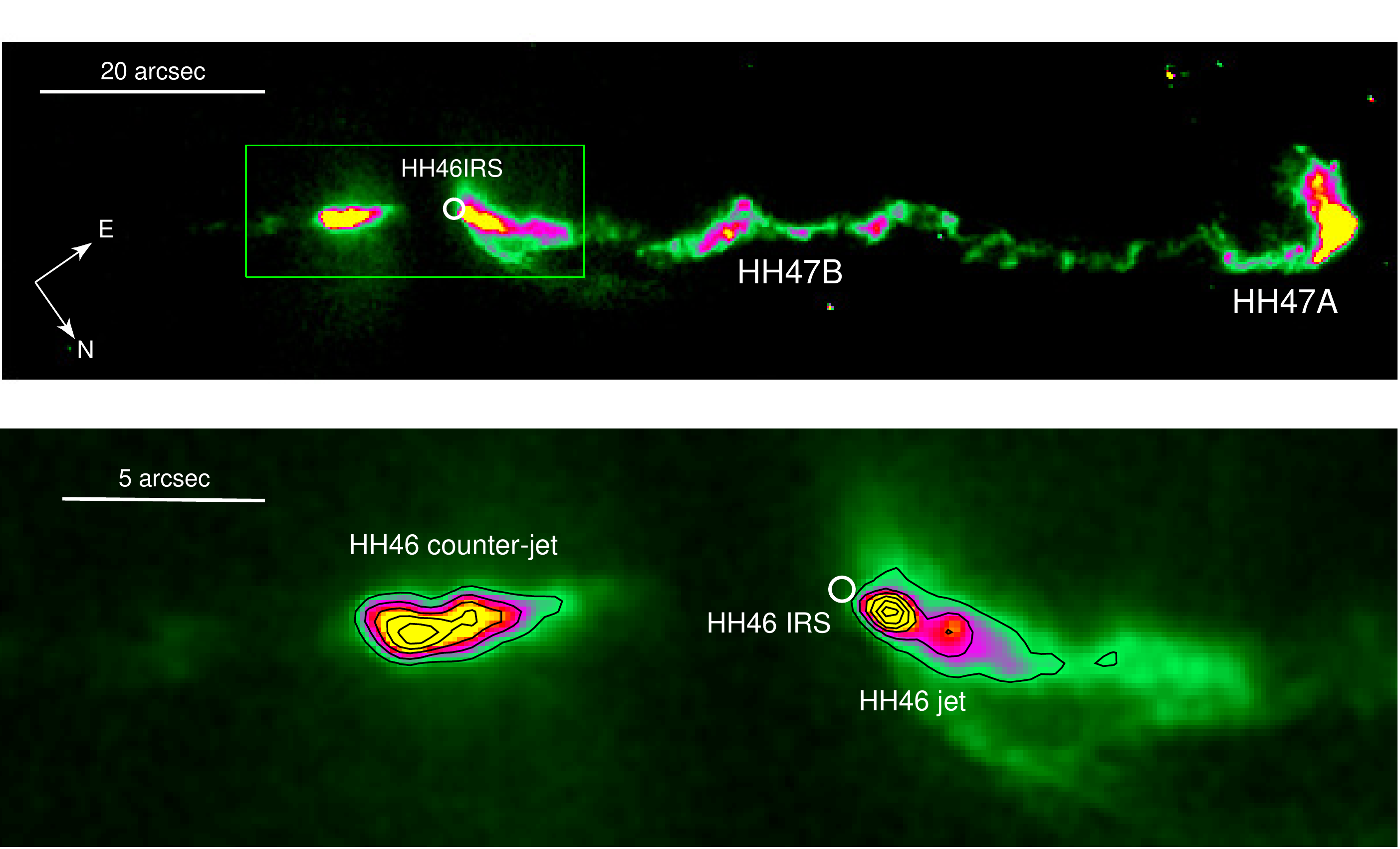}
    \caption{The same as Fig. \ref{fig:inner_regions_hh1} for the HH~46/47 region. Contour levels are at 1.3, 2.6, 4.9, 7.0, 10, and 15$\times$10$^{-19}$ \textbf{erg\,s$^{-1}$\,cm$^{-2}$\,\AA$^{-1}$\,pixel$^{-1}$.}}
    \label{fig:inner_regions_hh46}
\end{figure*}

\begin{figure*}[ht]
    \centering
    \includegraphics[width=0.8\textwidth,keepaspectratio]{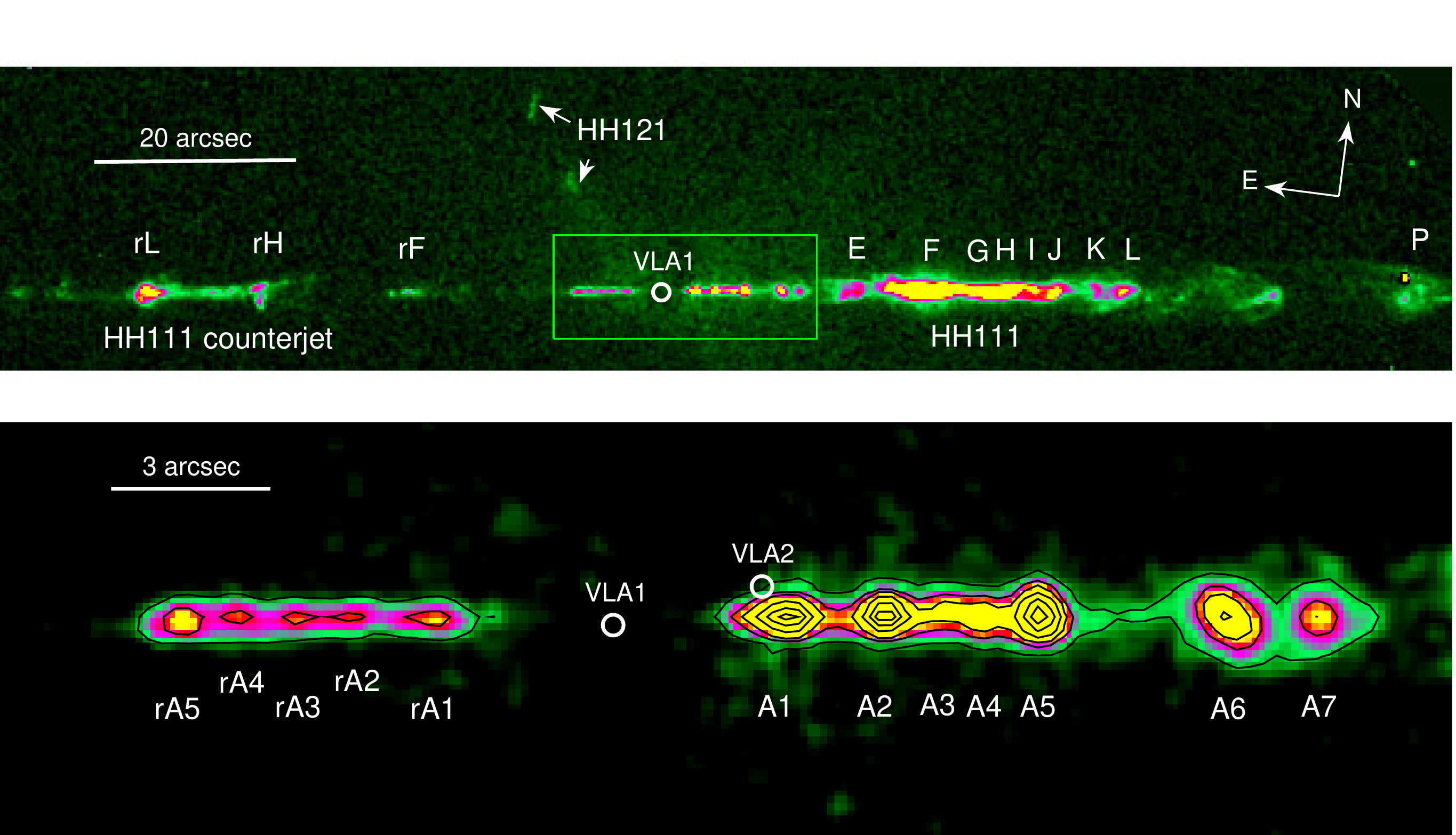}
    \caption{The same as Fig. \ref{fig:inner_regions_hh1} for the HH~111 region. Contour levels are at 0.2, 0.5, 0.8, 1.5, 2.5, 3.5 and 4.5$\times$10$^{-19}$ \textbf{erg\,s$^{-1}$\,cm$^{-2}$\,\AA$^{-1}$\,pixel$^{-1}$.}}
    \label{fig:inner_regions_hh111}
\end{figure*}

The top panel of Figures \ref{fig:inner_regions_hh1} to \ref{fig:inner_regions_hh111} gives the entire FoV, while the bottom panel provides an enlarged view of the region marked by the green box. A number of observed features are labelled in each figure.  

%\begin{itemize}
%    \item Figure 1 - pretty RGB composite with Fe (green), OI (blue) and continuum (red). Fe images to be coadded (1.25,1.64) and all images to be aligned,straightened etc and send to Pat
%    \item Include four figures, one for each jet, of the [FeII]1.64um image only, composed of panels with enlargement of  the emission in different jet regions (e.g. central region, bow shocks) - doing this in cont sub section?
%\end{itemize}

%\bru{I think we decided to remove the RGB images here and put instead the images in Fig.5, separated in 4 different figures and in a two column format.}

%\jess{JE: need to work on this part more}

%The counter-jet of HH~46 has previously been observed in the mid-IR \citep{NoriegaCrespo2004} and in the optical \citep{Hartigan2011} but it appears much fainter than in our images. The HH~111 counterjet has also been observed in the infrared (\citet{Reipurth1999}; \citet{NoreigaCrespo2011}). 

Figure \ref{fig:inner_regions_hh1} shows the image of the HH~1 and HH~2 region, that includes the HH~1 (blue-shifted) jet, driven by the radio source VLA1. The other radio source of the system, VLA2, driving the HH~144 outflow, is also indicated. 
Few weak knots of the counter-jet are also observed here. %for the first time. 
In the bottom panel, a region covering the HH~1 jet is enlarged. Here we see that the jet is detected as close as 2$\farcs$5 from the driving source while the innermost region is obscured due to high extinction, consistent with previous reports by \citet{Reipurth2000a, Davis2000hh1, Nisini2005}. Various jet knots are labelled from L to F following the nomenclature of \citet{Reipurth2000a}. The HH~501 knots, which differ in orientation with respect to the main HH~1 jet knots, are also labelled. The two knots in the red-shifted HH~1 counter-jet, correspond to similar distances from the source as knots G and F, and so are here labelled as rG and rF. 

%Here we see a number of knots in the inner jet region, two faint counterjet knots and an intersecting jet (HH~501). 

Figure \ref{fig:inner_regions_hh34} shows the HH~34 jet and associated HH34S bow shock to the south. The red-shifted counter-jet is clearly detected, at variance with optical and even [\ion{Fe}{2}] 1.25$\mu$m images. \citet{Stapelfeldt1991} first imaged the HH~34 jet in the NIR [\ion{Fe}{2}] 1.64$\mu$m line, and also faintly detected the counterjet (see Figure 7f in their paper). The counterjet was detected in further IR spectroscopic and imaging observations \citep{GarciaLopez2010, Antoniucci2014} and Spitzer images \citep{Raga2011}, but never with the level of details shown here. 
The bottom panel shows the various observed jet knots, which are labelled following the nomenclature of \citet{Reipurth2002}. The jet driving source, VLA, is also marked. 

Figure \ref{fig:inner_regions_hh46} shows the complex structure of the HH~46/47 target, which was first observed in the NIR by \citet{Eisloffel1994}. Our image covers most of the blue-shifted outflow, that includes the HH~46 jet and its red-shifted counter-jet, the HH~47B knots chain, and the bright bow shock towards the North-East, HH~47A. The position of the jet driving source, HH~46~IRS, is also marked. The enlarged figure of the continuum subtracted central region shows the detailed structure of the inner jet better than previous optical and IR \textit{HST} images as these were dominated by the strong reflection nebula in which the jet is embedded \citep{Reipurth2000a, Hartigan2005}. In fact, significant residual scattered line emission is still visible in the image on the northern side of the jet. Bright jet knots are seen that follow the arc-shaped structure of the jet. The red-shifted jet does not emerge until about 5 arcsec from the central source and it is not symmetrically displaced with respect to the blue-shifted jet. 

\begin{figure}[ht]
    \centering
    \includegraphics[width=\columnwidth,keepaspectratio]{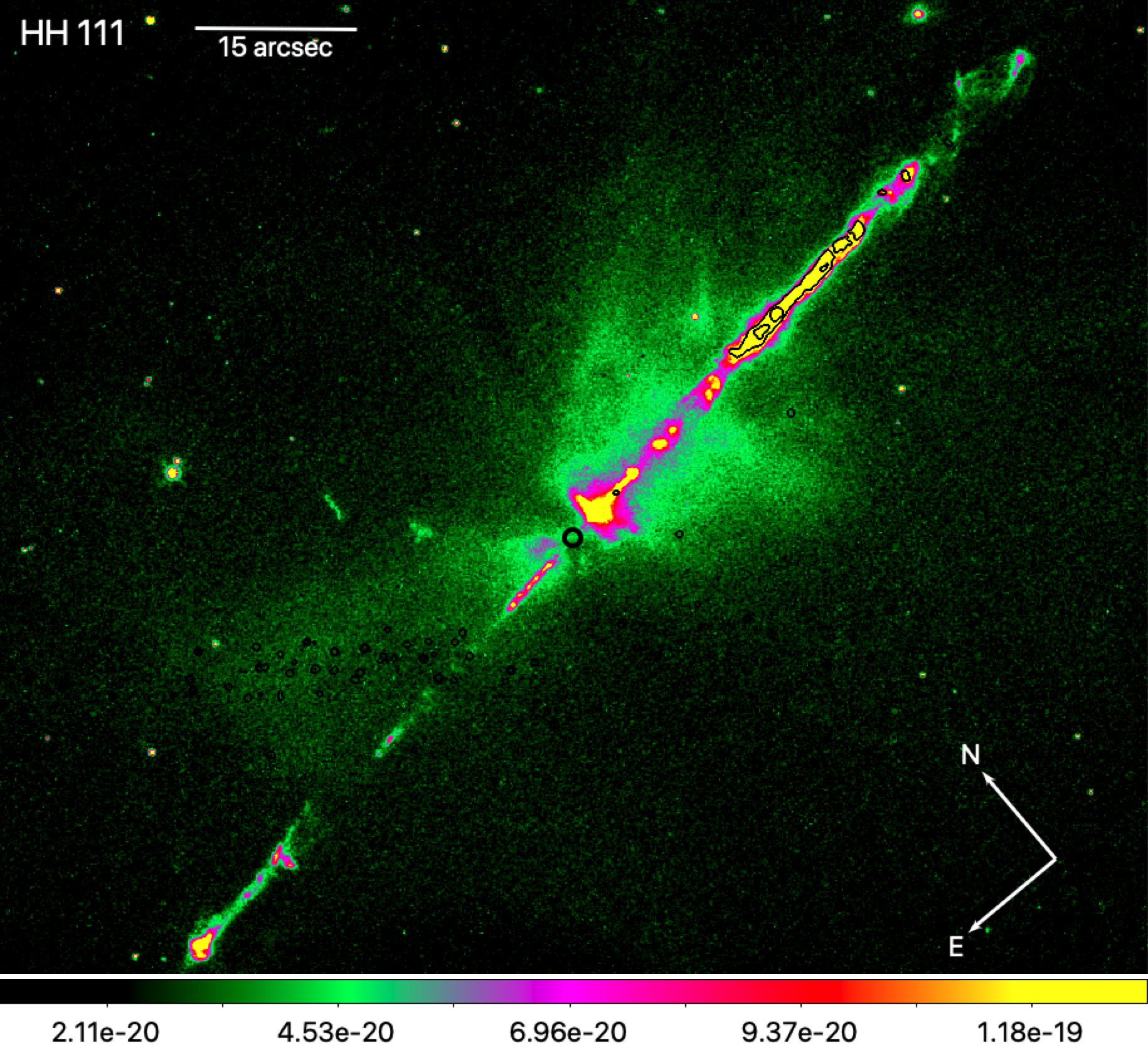}
    \caption{[\ion{Fe}{2}] 1.64$\mu$m line+continuum image of the HH~111 jet and cavity. [\ion{O}{1}] contours are overplotted in black (levels are 0.2 and 1 $\times$10$^{-19}$ \textbf{erg\,s$^{-1}$\,cm$^{-2}$\,\AA$^{-1}$\,pixel$^{-1}$}), showing that at optical wavelengths the jet is observed only at large distances from the source (marked by the black circle). The red-shifted jet is not detected in the optical. }
    \label{fig:HH111_cavity}
\end{figure}

Figure \ref{fig:inner_regions_hh111} shows the HH~111 bipolar jet, driven by the radio source VLA1 with the bright blue-shifted lobe and its dimmer red-shifted counterpart. Many individual knots are revealed in the enlarged view of the inner jet region. %Almost perpendicular to 
Near the HH~111 jet we observe also knots from the HH~121 jet, driven by the VLA2 radio source.
The [\ion{Fe}{2}] 1.64$\mu$m image reveals many details of the jet structure within $\sim$ 20 arcsec from the driving source (VLA1) that remained hidden in optical images, where the jet is seen only at larger distances \citep[knots from E to L,][]{Reipurth1997,Hartigan2001}, when it emerges from a cone-shaped cavity. This is shown in Fig. \ref{fig:HH111_cavity}, where the image taken in the F167 continuum filter is presented, with superimposed contours of the [\ion{O}{1}] emission. 
%\bru{Here we could show the cone-shaped cavity with a NIR continuum image like the one I have provisionally included here, with superimposed OI contours}
Some of the red-shifted knots observed here were already discovered in previous {\it HST}/NICMOS and Spitzer images of the jet \citep{Reipurth2000a,NoreigaCrespo2011} as well as in ground-based 2.12$\mu$m observations \citep{Coppin1998}. However, our continuum subtracted image, where the large nebulosity around the central source is removed, reveals the sequence of symmetric blue- and red-shifted knots with an unprecedented level of detail. We name the inner knots, observed only in the IR in the blue-shifted jet, A1 - A7. We name corresponding counter-jet knots rA1 - rA5.  

\subsubsection{Proper Motions}
\label{section:time_var}
%\begin{itemize}
%    \item Use Pat's SII and H$\alpha$ (2-3 epochs) to look for new features and new knot positions with time. Data to be realigned using same method as above (ccmap, ccxymatch).
%    \item Using velocities in Hartigan et al 2001, can estimate where the knots should be positioned now, and if there's a difference from this - discuss.
%    \item HH34 shows a few newhttps://www.overleaf.com/project/5d9b09231528080001a70563 features/blobs brighter than before etc (looked at in Rome)
%    \item Figures like Fig 3 of Antoniucci et al 2014
%\end{itemize}

% haven't done HH1 yet since there weren't enough stars in the images but can change this to "all sources" etc when/if HH1 is done

%We present fourth epoch [\ion{Fe}{2}] images of our chosen targets (third for HH~111).

\begin{figure*}
    \centering
    \includegraphics[width=0.7\textwidth,keepaspectratio]{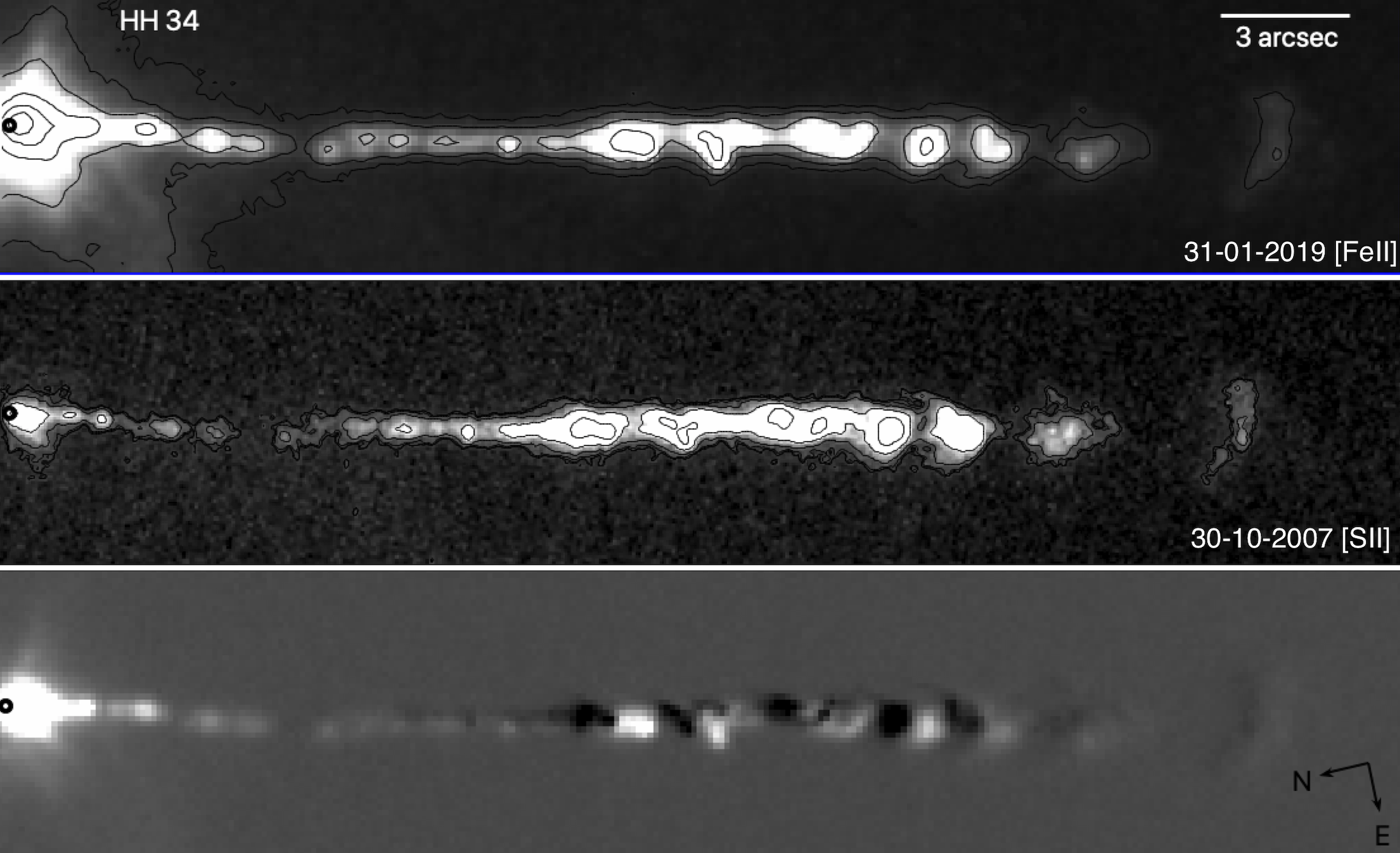}
    \caption{Comparison of the HH~34 knot positions in [\ion{Fe}{2}] with contour levels at 0.05, 0.08, 0.19, 0.67, 2.5 and 9.9  $\times$10$^{-18}$ \textbf{erg\,s$^{-1}$\,cm$^{-2}$\,\AA$^{-1}$\,pixel$^{-1}$.} (top) and [\ion{S}{2}] with contour levels at 0.6, 0.63, 0.74, 1.2, 2.9 and 9.9  $\times$10$^{-16}$ \textbf{erg\,s$^{-1}$\,cm$^{-2}$\,\AA$^{-1}$\,pixel$^{-1}$.} (middle) images with an eleven year baseline. %\bru{Check is the eleven year baseline is the same for all the jets. Include within the images the date of observation}.
    The circle marks the driving source position. The bottom panel shows the subtraction image for [\ion{Fe}{2}]-[\ion{S}{2}]}
    \label{fig:hh34_pm}
\end{figure*}

\begin{figure*}
    \centering
    \includegraphics[width=0.6\textwidth,keepaspectratio]{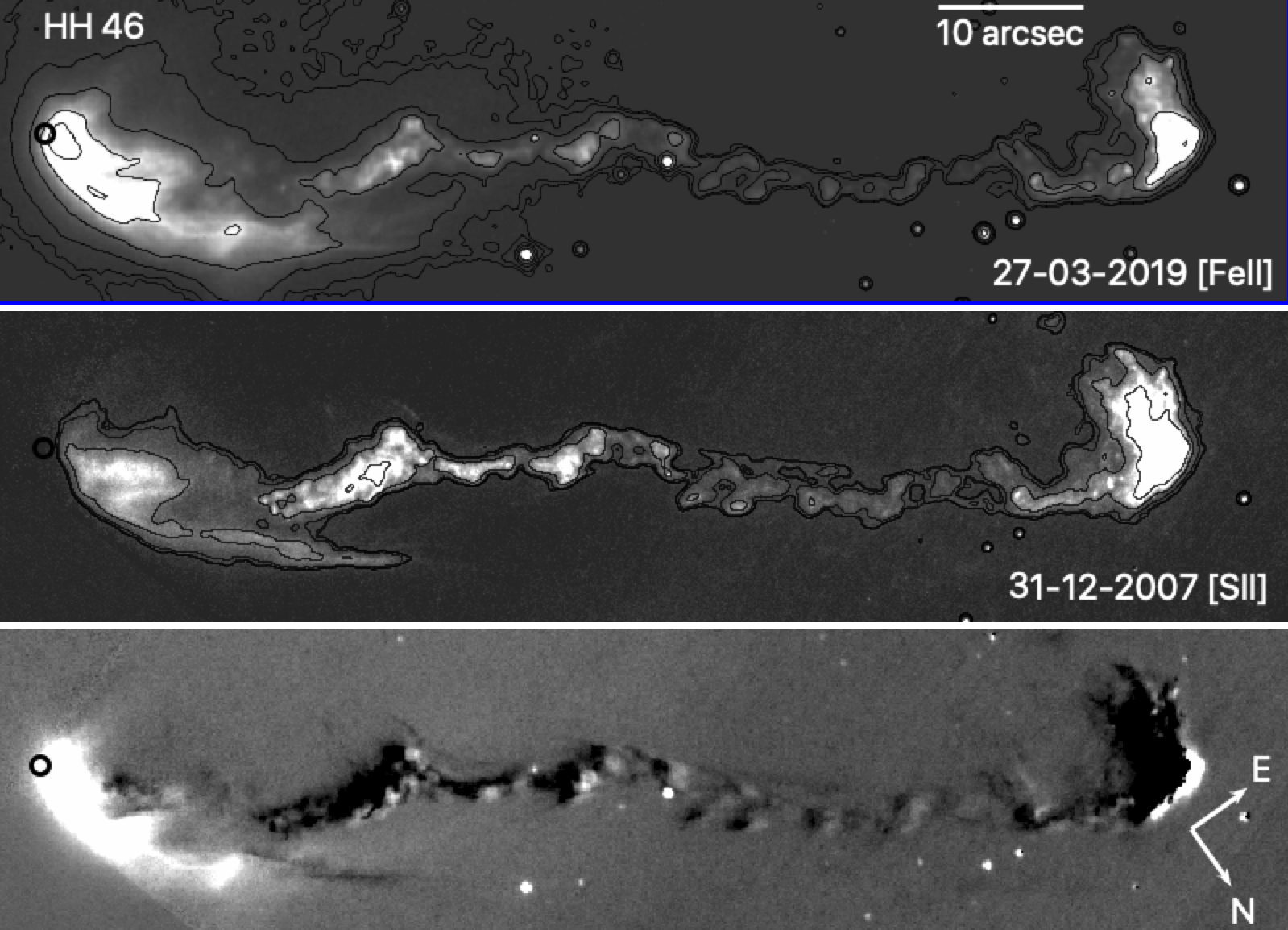}
    \caption{Same as Fig. \ref{fig:hh34_pm} for the HH~46 outflow. [\ion{Fe}{2}] contour levels are 0.2, 0.23, 0.35, 0.8, 2.6 and 9.9 $\times$10$^{-19}$ \textbf{erg\,s$^{-1}$\,cm$^{-2}$\,\AA$^{-1}$\,pixel$^{-1}$}. [\ion{S}{2}] contour levels are 0.9, 0.94, 1.1, 2.5 and 9.9 $\times$10$^{-16}$ \textbf{erg\,s$^{-1}$\,cm$^{-2}$\,\AA$^{-1}$\,pixel$^{-1}$.}}
%    Comparison of the HH~46 knot positions in [\ion{Fe}{2}] (top) and [\ion{S}{2}] (middle) images with an eleven year baseline. Xs mark the brightest point in each knot. The circle marks the source position. The bottom panel shows the subtraction image for [\ion{Fe}{2}]-[\ion{S}{2}].}
    \label{fig:hh46_pm}
\end{figure*}

\begin{figure*}
    \centering
    \includegraphics[width=0.7\textwidth,keepaspectratio]{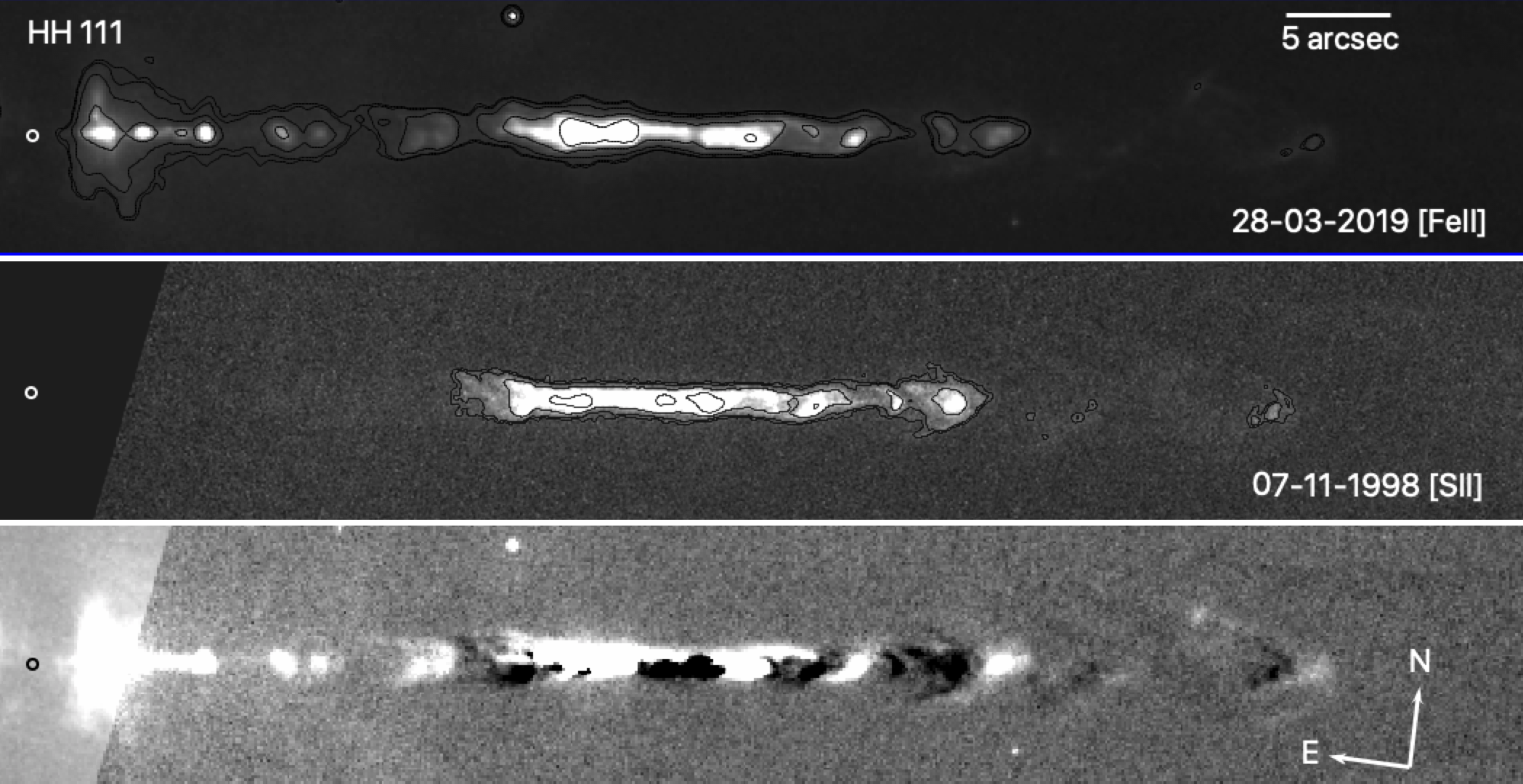}
    \caption{Same as Fig. \ref{fig:hh34_pm} for the HH~111 outflow. [\ion{Fe}{2}] contour levels are 0.55, 0.58, 0.75, 1.7 and 7 $\times$10$^{-19}$ \textbf{erg\,s$^{-1}$\,cm$^{-2}$\,\AA$^{-1}$\,pixel$^{-1}$.} [\ion{S}{2}] contour levels are 0.1, 0.12, 0.25, 0.9 and 5 $\times$10$^{-15}$ \textbf{erg\,s$^{-1}$\,cm$^{-2}$\,\AA$^{-1}$\,pixel$^{-1}$.}}
%    Comparison of the HH~111 knot positions in [\ion{Fe}{2}] (top) and [\ion{S}{2}] (middle) images with an eleven year baseline. Xs mark the brightest point in each knot. The circle marks the source position. The bottom panel shows the subtraction image for [\ion{Fe}{2}]-[\ion{S}{2}].}
    \label{fig:hh111_pm}
\end{figure*}

We compare the [\ion{Fe}{2}] images of our targets with archival images taken in previous epochs with \textit{HST} in the [\ion{S}{2}] filter (top two panels of Figures \ref{fig:hh34_pm}-\ref{fig:hh111_pm}). The [\ion{S}{2}] images of the HH~34 and HH~46 jets used here were first published in \citet{Hartigan2011}; and for HH~111 we use the [\ion{S}{2}] image from \citet{Hartigan2001}.
%\bru{We should probably cite here the papers were these images have been published for the first time. Knowing the observing date, it should not be difficult to retrieve this information.}

%The top two panels in Figures \ref{fig:hh34_pm}-\ref{fig:hh111_pm} show the [\ion{Fe}{2}] and [\ion{S}{2}] images. 

The jet morphology, when both lines are detected, is very similar with only minor changes in individual knots. 
The major difference between the two tracers is that [\ion{Fe}{2}] emission is observed close to the source, unlike [\ion{S}{2}]. This is particularly evident for HH~111 (see Figure \ref{fig:hh111_pm}), where [\ion{S}{2}] is observed only where the jet emerges from the cone-like cavity. 

As the [\ion{S}{2}] images were taken typically 10-20 years before the [\ion{Fe}{2}] images, combining the two datasets allows us to measure the secular proper motions of the jets with great accuracy, as although the tracers are different, the [\ion{S}{2}] and [\ion{Fe}{2}] emission is expected to peak in the same post-shock region for a given epoch \citep{Nisini2005}.
%Note that [\ion{Fe}{2}] 1.64~$\mu$m and [\ion{S}{2}]$\lambda$6731 trace similar post-shock regions \cite{Nisini2005}, and so it is appropriate to compare the two tracers to measure proper motions between epochs. 
%The proper motion of HH1 could not be performed, since no stars were found on the SII optical image for a proper orientation and alignment with the %FeII image.

Firstly, the [\ion{Fe}{2}] and [\ion{S}{2}] images were registered to each other using field stars detected in both filters. For HH~1 however, only one star in common was found in the two fields, therefore it was not possible to align the images. We therefore report on the proper motion for the HH~34, HH~46/47 and HH~111 jets only. 
The images were also rotated such that the direction of the flow was aligned to the horizontal according to the jet PAs listed in Table \ref{table:targets}, and defined on the inner jet/counter-jet knots.
%(31.4$^{\circ}$ and 141.9$^{\circ}$ counter-clockwise for HH~34 and HH~46, respectively; and 47.4$^{\circ}$ clockwise for HH~111)
%\bru{these numbers need to be consistent with the PA given in table 1} \jess{these are measured in image coordinates, so they're different from the jet PA measured from N to E}. 
The proper motions were derived by measuring the shifts between the individual knots in the [\ion{Fe}{2}] and [\ion{S}{2}] images. These shifts are clearly observed in the [\ion{Fe}{2}] - [\ion{S}{2}] difference images for each target as presented in the bottom panels of Figures \ref{fig:hh34_pm}-\ref{fig:hh111_pm}. We identified knots that do not present major structural changes between one epoch and the next, and we measure the difference in position between the photo-centres of the knots in the two epochs. Proper motions converted to tangential velocities using the adopted distance, together with values reported in recent literature based on a smaller time interval corrected for our adopted distance, are presented in Figure \ref{fig:velocities_all} and listed in the tables of Appendix~\ref{appendix_b}
%Therefore, we overplot previously measured velocities in Figures \ref{fig:hh34_pm}-\ref{fig:hh111_pm} (and give details in the tables of Appendix~\ref{appendix_b}), correcting for our adopted distance when necessary. 

For the HH~34 jet, the measured tangential velocities (top panel of Figure \ref{fig:velocities_all}) tend to decrease with distance from the source, as already reported by \citet{Raga2012}, with differences of within 20-30~\kms\ in most of the knots. 
%The velocities from \citet{Raga2012} were corrected to a distance of 383~pc and plotted in black in Figure \ref{fig:velocities_all}. 
Ground based images of the HH~34 jet \citep{Eisloeffel1992} report a similar trend for the inner part of the jet, however a large increase in tangential velocity was measured further from the source which is not seen in our results.
At the position of about 5$\arcsec$, we see an abrupt decrease of tangential velocity which is not recorded by Raga et al. However, the same rapid decrease  at similar distances is measured in previous proper motion studies by \citet{Eisloeffel1992} and \citet{Reipurth2002}. 

For the HH~111 jet, the tangential velocities decrease with distance in the first 40$\arcsec$ from the star, with small fluctuations of about 15-20 \kms. This trend was also found by \citet{Hartigan2001}, with higher absolute values by approximately 20-30~\kms. The velocity seems then to increase again at further distance from the star. 

The deceleration of jets with distance is a known phenomenon observed in several extended outflows. It has been interpreted as either due to an intrinsic variability of the ejection velocity or to a braking in the interaction of the jet with the surrounding medium. Jet precession was also suggested as a possible cause for the jet slow down \citep[e.g.][]{Masciadri2002}. 

At variance with the other two outflows, the HH~46/47 jet appears to be accelerating with distance from the star.
If we exclude the knots at 40$\arcsec$ and 70$\arcsec$, the derived tangential velocities correspond to those found by \citet{Hartigan2005} within 20~\kms. Ground based images \citep{Eisloffel1994pm} measure similar tangential velocities which generally remain more constant along the jet length.

\begin{figure*}
    \centering
    \includegraphics[width=0.7\textwidth,keepaspectratio]{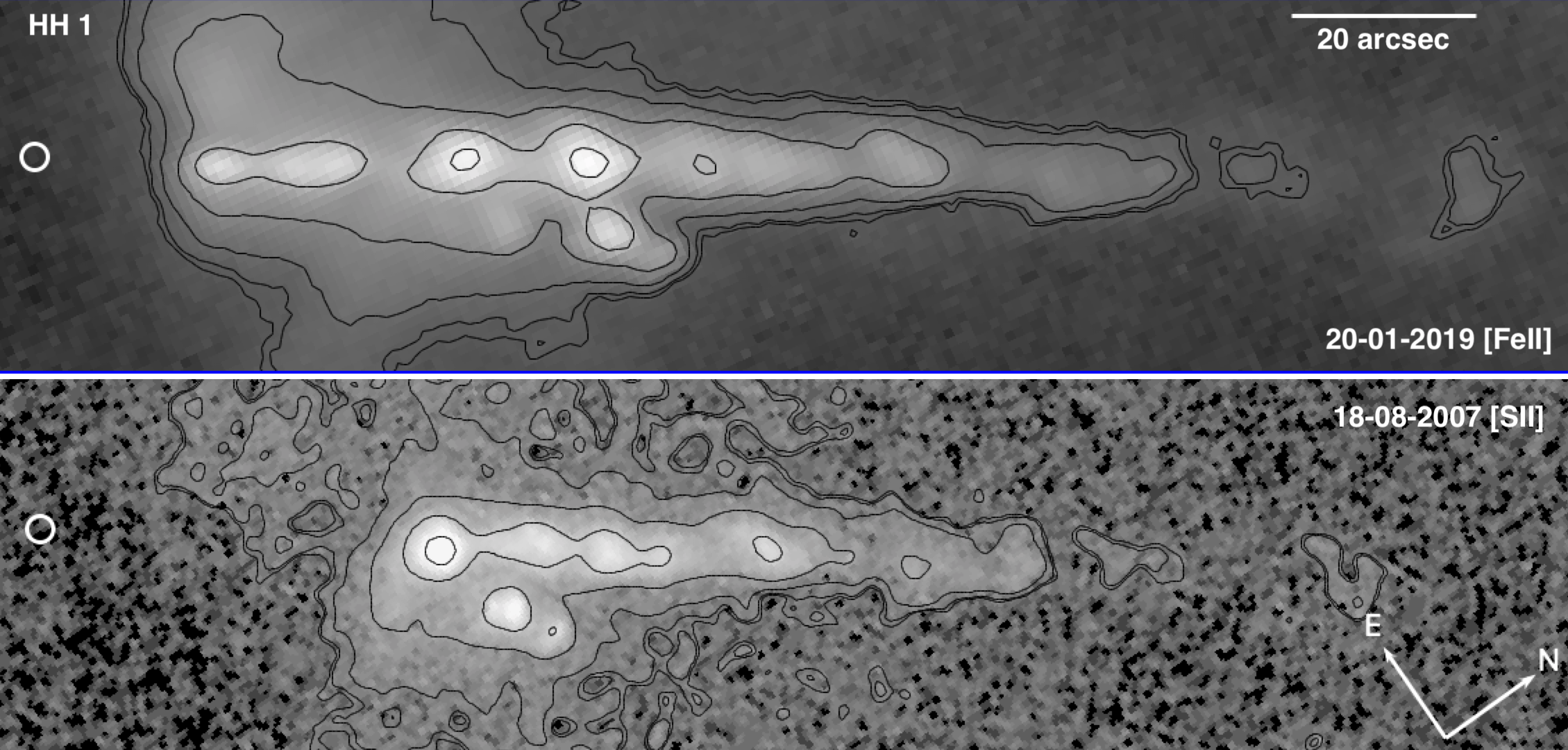}
    \caption{Comparison of HH~1 [\ion{Fe}{2}] (top) and [\ion{S}{2}] (bottom) images with an eleven year baseline. Only one star is present in both images so it was not possible to accurately align the images to create the subtraction image. However, we still see the jet in the  [\ion{Fe}{2}] image is detected much closer to the star than in [\ion{S}{2}]. [\ion{Fe}{2}] contour levels are 0.07, 0.08, 0.1, 0.25, 0.8 and 3 $\times$10$^{-18}$ \textbf{erg\,s$^{-1}$\,cm$^{-2}$\,\AA$^{-1}$\,pixel$^{-1}$.} [\ion{S}{2}] contour levels are 0.4, 0.42, 0.5, 0.8, 2 and 7 $\times$10$^{-15}$ \textbf{erg\,s$^{-1}$\,cm$^{-2}$\,\AA$^{-1}$\,pixel$^{-1}$.}}
%    Comparison of the HH~111 knot positions in [\ion{Fe}{2}] (top) and [\ion{S}{2}] (middle) images with an eleven year baseline. Xs mark the brightest point in each knot. The circle marks the source position. The bottom panel shows the subtraction image for [\ion{Fe}{2}]-[\ion{S}{2}].}
    \label{fig:hh1_pm}
\end{figure*}

\begin{figure}
    \centering
    \includegraphics[width=0.4\textwidth]{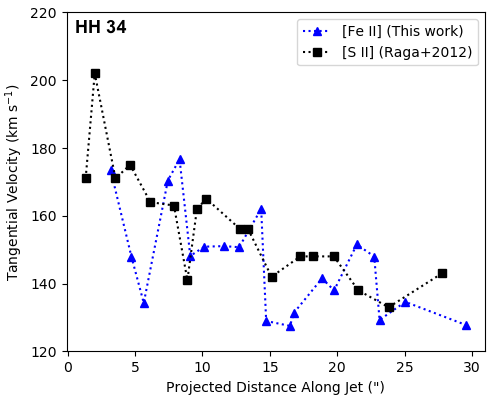}
    \includegraphics[width=0.4\textwidth]{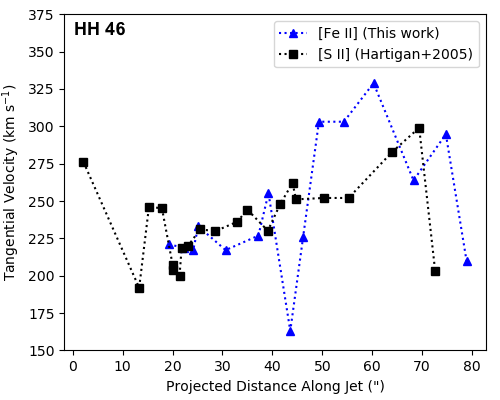}
    \includegraphics[width=0.4\textwidth]{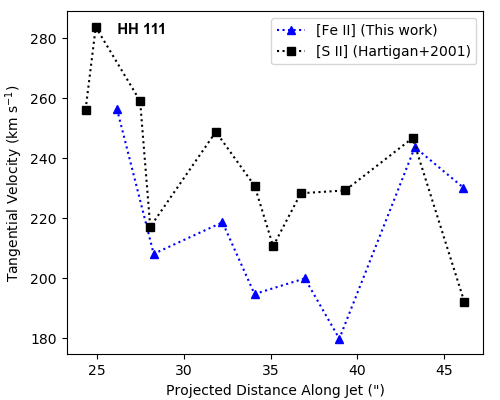}
    \caption{Tangential velocities for HH~34 (top), HH~46 (middle) and HH~111 (bottom). The black markers show literature values, while the blue markers are the values found from the [\ion{Fe}{2}] knot positions in this work.}
    \label{fig:velocities_all}
\end{figure}

\subsection{Inner Jet Region}

\textit{HST} infrared images allow us to trace the jet targets closer to the central source than was previously possible with optical images, and to detect the fainter counter-jet in all cases. These images can therefore provide insights into some properties of the inner jet region which have so far remained unexplored. In this section we analyse the width of the jets and its variation with distance from the central source, the symmetry between the jet and the counter-jet, and the extinction in the inner jet region. 
%\subsubsection{Continuum-subtracted images}

%Continuum subtraction for each line – line-continuum. Some continuum images scaled, some over subtraction/nebulosity remaining in output files. \jess{I think we decided not to scale the continuum images to reduce any noise added because of CS?}

%\begin{itemize}
%    \item Continuum subtraction procedure i.e. after astrometry and photometry, do scaling of continuum and subtract
%    \item Continuum subtracted images with boxes inset for close ups of inner region close to the source \bru{I think the images to present here are only the close-up images that I have already included. At larger scale we do not expect that the continuum subtracted images will be very different}
 %   \item Weaker excess emission lines and line scattering in continuum image introducing errors in continuum subtracted images and ratio maps. Include here a figure showing the filter response and the location of the emission lines \bru{I still think this is important and we can make this figure or a table, like I suggested before. This last point should be addressed before}
%\end{itemize}
% FIGURES OF THE INNER JET REGION
%

%%% Dunham et al 2014 "SEDs of Class 0 protostars are dominated by the far-infrared component and the near-infrared is generally dominated by scattered light from the outflow cavity"

\subsubsection{Jet Collimation}
\label{section:width}

%Jet width – 1.64~$\mu$m images rotated so jet is aligned with horizontal x-axis. 1D gaussian fitted column-by-column, FWHM of fit taken as jet width. Filtered to only plot columns where fit amplitude > 0 (to avoid fitting to negative noise peaks). 

The degree of collimation of the jets can be estimated by measuring how the jet width varies with distance from the driving source.
In order to measure the jet width, the continuum-subtracted [\ion{Fe}{2}] 1.64~$\mu$m images were first rotated to align the jet PA with the x-axis (see Section \ref{section:time_var}). 
Two approaches were then considered. In the first, we measured the jet width as the FWHM of the transverse intensity profile with a single or double Gaussian, depending on the presence of low-intensity wings. The second approach, was to directly measure the width of the transverse intensity profile at half the height of the peak, without Gaussian fitting. The two methods give similar results within the errors. In some cases, as in HH~1 at the position where the HH~501 knots intersect the main jet, a triple-Gaussian fit was required.
%to accurately reproduce the combined contributions of the jet emission, the HH~501 emission and the remnants of emission from the circumstellar nebulosity.
Lastly, the jet widths were deconvolved by subtracting in quadrature the instrumental FWHM of 0$\farcs$153. 

%In the case of HH~34, HH~47 and HH~111, the jet width was measured by fitting the intensity profile with a single gaussian, and considering the corresponding FWHM for each x-position along the jet. In some cases, where an additional low-intensity component was present, the profile was fitted with a double gaussian to take into account the extended wings. A comparison of the fitted profiles for a bright and dim position along the jet is presented in Figure \ref{fig:compare_profiles}. For HH~1, the presence of the intersecting jet made it impossible for a single 1D Gaussian fit to accurately determine the jet width (see Figure \ref{fig:HH1_with_501}). Instead, a triple-Gaussian fit was required to accurately fit the jet, intersecting jet from a different source and the remnants of emission from the circumstellar nebulosity. The jet widths were deconvolved by subtracting in quadrature the instrumental FWHM of 0$\farcs$153.

%\begin{figure}
%    \centering
%    \includegraphics[width=0.9\columnwidth]{HH1_w501.jpg}
%    \caption{Comparison of the brightness profiles for two sample positions along the HH~1 jet. The blue curve represents a part of the jet where it is intersected by HH~501, which makes it impossible to measure the jet width using a single 1D Gaussian.}
%    \label{fig:HH1_with_501}
%\end{figure}

\begin{figure*}
   \centering
    \includegraphics[width=0.99\columnwidth]{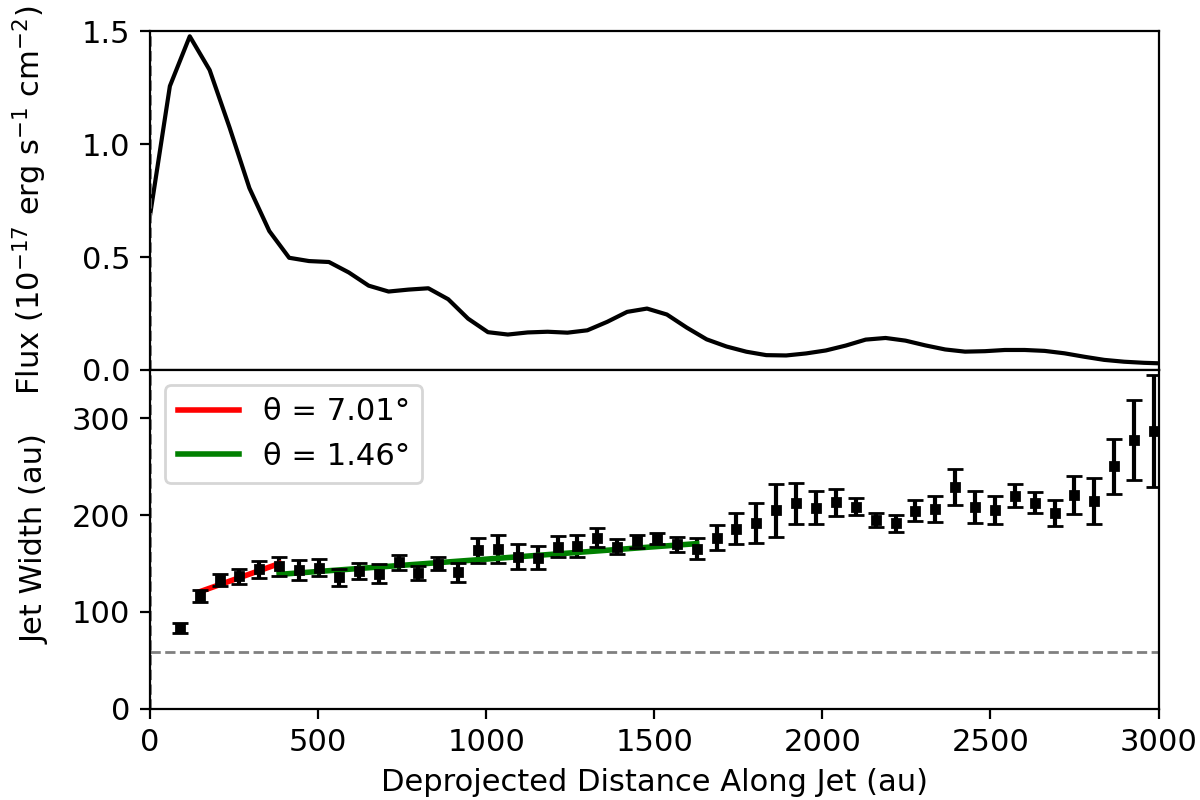}
    \includegraphics[width=0.99\columnwidth]{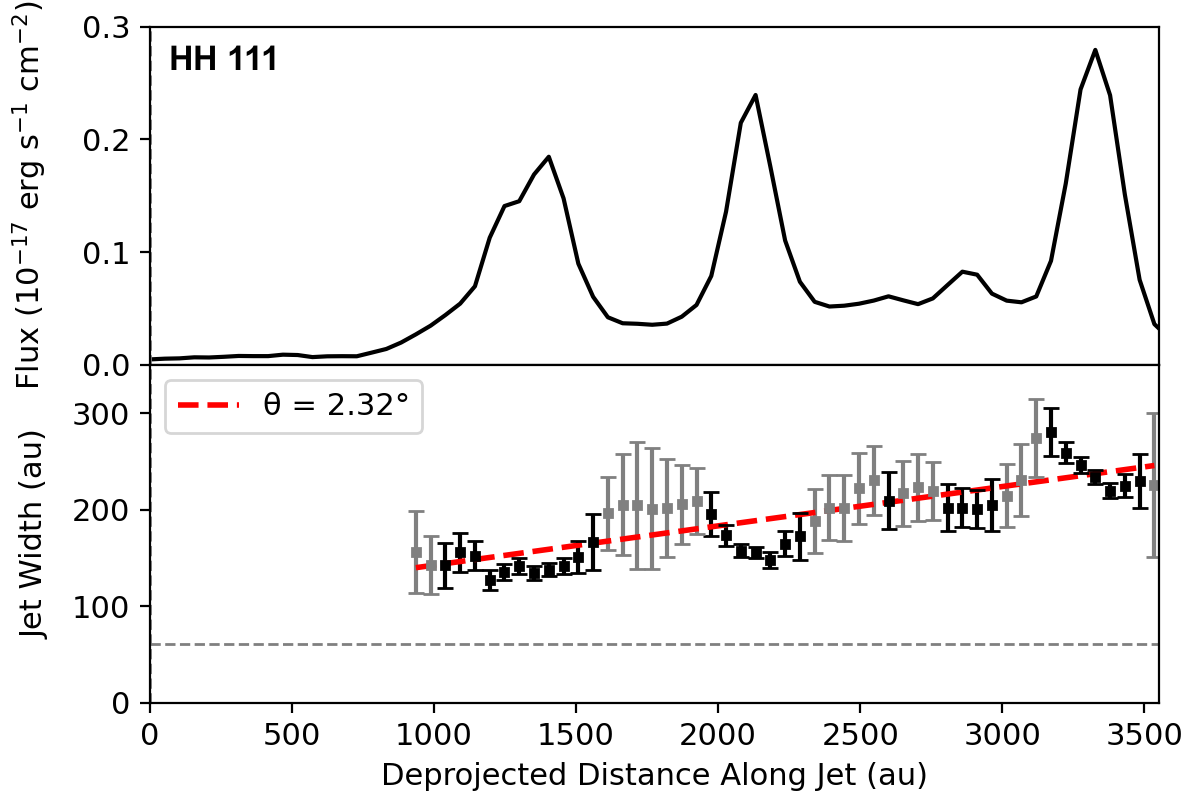}%%
    \\    
    \includegraphics[width=0.99\columnwidth]{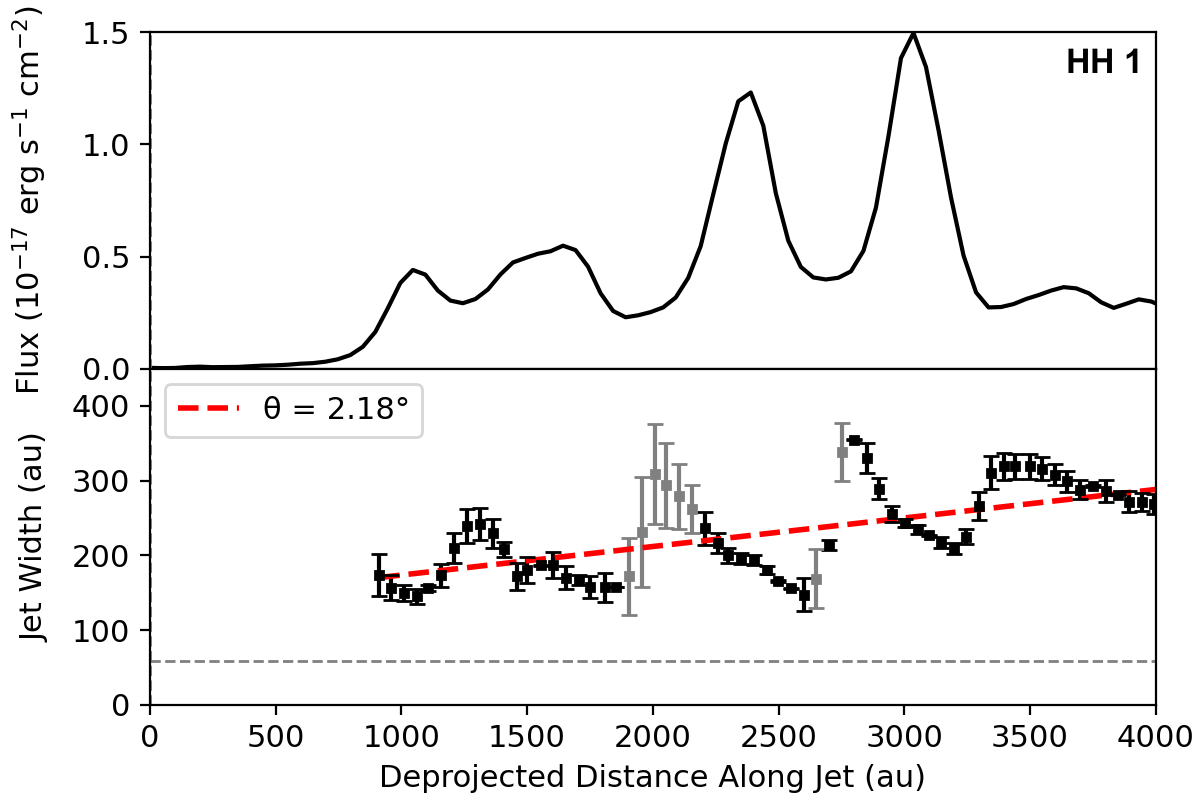}
    \includegraphics[width=0.99\columnwidth]{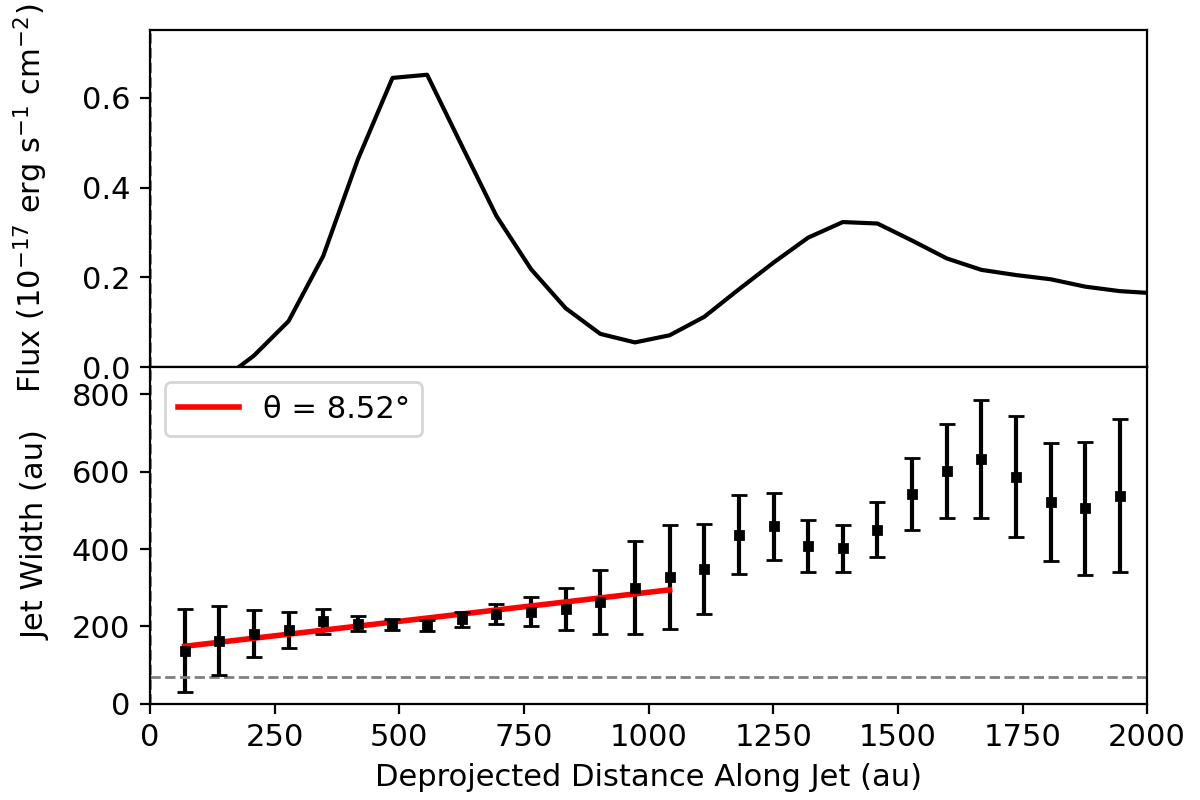}%%
    \caption{Top: Binned flux along the jet. Bottom: Jet FWHM from 1D Gaussian fits at each pixel. Errors are 3$\sigma$. In HH~34, we measure the opening angle with a linear fit over the regions in the jet marked by red and green solid lines. In HH~46, the opening angle is measured at every point along the jet up to approx 1000~au, marked by the solid red line. For HH~1 and HH~111, the opening angle is measured by fitting all points with a 3$\sigma$ smaller than 30~au (dashed red lines). The grey points in for HH~1 and HH~111 are those with 3$\sigma$ larger than 30~au, which are not fitted when measuring the opening angle. The dashed grey line is the instrumental FWHM of 0$\farcs$153.}
    \label{fig:jfwhm_all}
\end{figure*}

Figure \ref{fig:jfwhm_all} shows the measurements of jet widths for each of the targets as a function of distance from the central source. For comparison, the binned flux along the jet is plotted in the top panels. 
%Previous measurements of the jet widths were taken on a few knots along the jet (e.g. \citet{Hartigan2011}). Here, we observe the jet at infrared wavelengths allowing us to better trace the inner regions of the jet, in addition to the high spatial resolution of HST.
%\bru{Discuss here comparison with previous measurements and highlight better sampling and better tracing of the inner regions in our data.}
In all cases, a gradual increase in jet width with distance is found, with an undulation in the jet width for the HH~1 and HH~111 jets, anticorrelated with the intensity. 
%The broader widths appear to correspond to dimmer points along the jet.
This effect is due to the fact that in the dimmer points the residual nebulosity dominates the emission and consequently the jet FWHM is overestimated.
%The fitting routine appears to overestimate the FWHM at the dimmer points, due to the more dominant presence of the residual nebulosity (see Figure \ref{fig:compare_profiles}). 
This undulation is not observed at HH~34 or HH~46. 
In all cases, the jet width could not be measured closer than $\sim$100~au to the star either because of a lack of jet emission, or residual continuum emission close to the star. 
%The jet widths were measured only for the inner knots, because in the outer regions it is not possible to accurately determine the FWHM as the material collides in shocks and/or interacts with the surrounding material. 

\begin{figure}
    \centering
    \includegraphics[width=0.9\columnwidth]{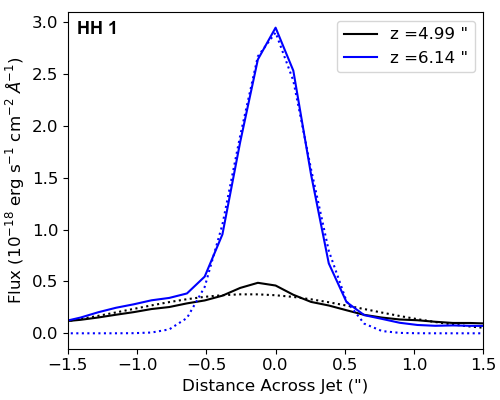}
    \includegraphics[width=0.9\columnwidth]{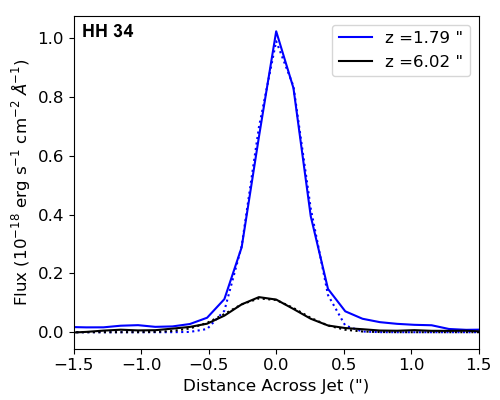}
    \caption{Top: Comparison of the brightness profile and 1D Gaussian fits for two sample positions along the HH~1 jet. The dim profile (black) shows how the jet width may be overestimated by the presence of low intensity emission, leading to the undulating pattern of the jet width seen in Figure \ref{fig:jfwhm_all}. Bottom: Comparison of the brightness profile and 1D Gaussian fits for two sample positions along the HH~34 jet which does not appear to be affected by residual low intensity emission.}
    \label{fig:compare_profiles}
\end{figure}

 The observed expansion of the de-projected jet width with distance gives a measure of the opening angle of the jet. As shown in Figure \ref{fig:jfwhm_all}, opening angles of $\approx$ 2$^{\circ}$ are found for the outer regions of the jets (i.e. at deprojected distances larger than a few hundred au from the source). However, for HH~46, we observe a wider opening angle of $\approx$ 8.5$^{\circ}$. Furthermore, for HH~34 we observe a change in opening angle, where in the innermost part of the jet we measure an opening angle of 7$^{\circ}$, decreasing to 1.5$^{\circ}$ measured between $\approx$ 400 - 1500 au from the star. Table \ref{table:angle} provides the values measured for each target.
 
 Estimates of the widths for individual knots of the HH~1, HH~34 and HH~111 jets were given from previous \textit{HST} images in \citet{Reipurth2000a}, \citet{Reipurth2002}, \citet{Reipurth2000b}, respectively. However these works, which are based on images without continuum subtraction of the source nebulosity, mainly addressed jet collimation at larger distances from the source. In \citet{Reipurth2000a, Reipurth2000b} the jet widths and opening angles of HH~111 and HH~1 were measured for images in both [\ion{S}{2}] and [\ion{Fe}{2}] emission, the latter being acquired with {\it HST}/NICMOS observations. In HH~111, these previous measurements give a width of about 150~au at a distance of $\sim$ 1200~au, i.e. compatible with our estimates. For HH~1, \citet{Reipurth2000b} derive a width of about 80~au at a distance of 1000~au, but we measure about 150~au.  Meanwhile, at larger distances (e.g. knot H and G) there is better  agreement. For the HH~34 jet, \citet{Reipurth2002} report a variation of the [\ion{S}{2}] width with distance. \textbf{Their measured widths are smaller than ours for the inner knots (at $<$ 2 arcsec from the source), while at larger distances the two become comparable. We think that these differences are caused by the increased difficulty in separating the jet from the source nebulosity in images where the continuum is not subtracted.} 
 
 %, due to the large nebulosity superimposed on the jet emission.  
 %These previous works pointed out that the jet widths were very similar in both the optical and IR tracers, as expected since [\ion{S}{2}] and [\ion{Fe}{2}] are excited under similar conditions. We confirm this result performing the width measurement also on the [\ion{S}{2}] images, finding a similar trend with distance, as can be seen in Fig. \ref{fig:fe_sii}.
 
 %\bru{I think that in the figure we should also plot the result Jessica made on the widths measured on the SII images, to show that they are in agreement with those of FeII and not with the Reipurth et al. results}
 
% \begin{figure}
%%%%% \end{figure}

\begin{table}[ht]
\centering
\caption{Opening angles for each target measured over various distances along the jet}
\begin{tabular}{|ccc|}
\hline
Target                 & z  & Opening angle \\
                 & (au)  & ($^{\circ}$) \\\hline
HH 1                   & \textgreater 1000 & 2.2                       \\
\hline
\multirow{2}{*}{HH 34} & \textless 400     & 7.0                       \\
                       & 400 - 1500  & 1.5                        \\
\hline
HH 46 & \textless 1000  & 8.5                        \\
\hline
HH 111                 & \textgreater 1000 & 2.3                        \\ \hline
\end{tabular}
\label{table:angle}
\end{table}

\subsubsection{Jet/counter-jet asymmetries} \label{sec:jaxis}

\begin{figure*}
    \centering
    \includegraphics[width=0.95\columnwidth]{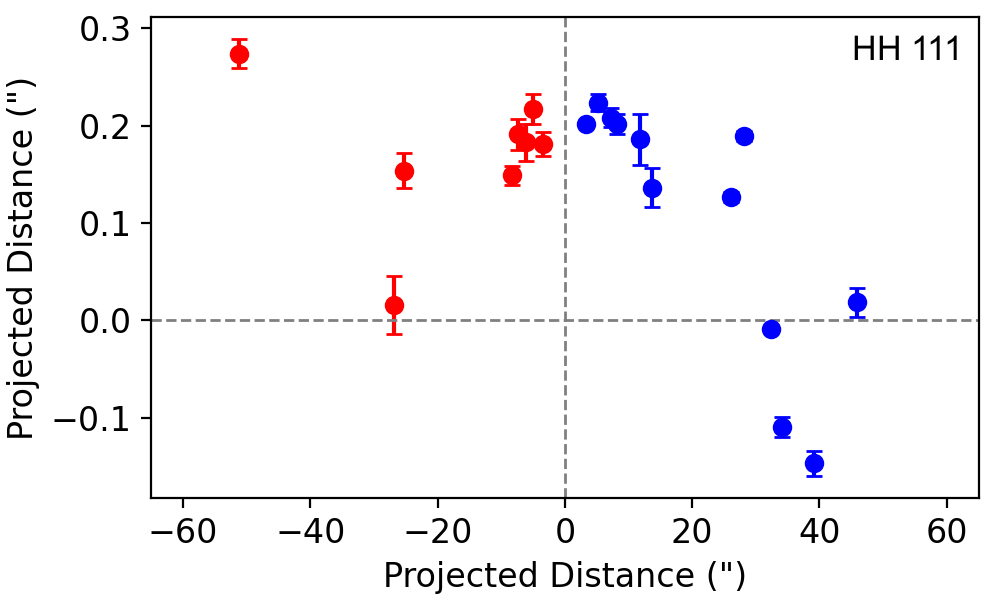}
     \includegraphics[width=0.95\columnwidth]{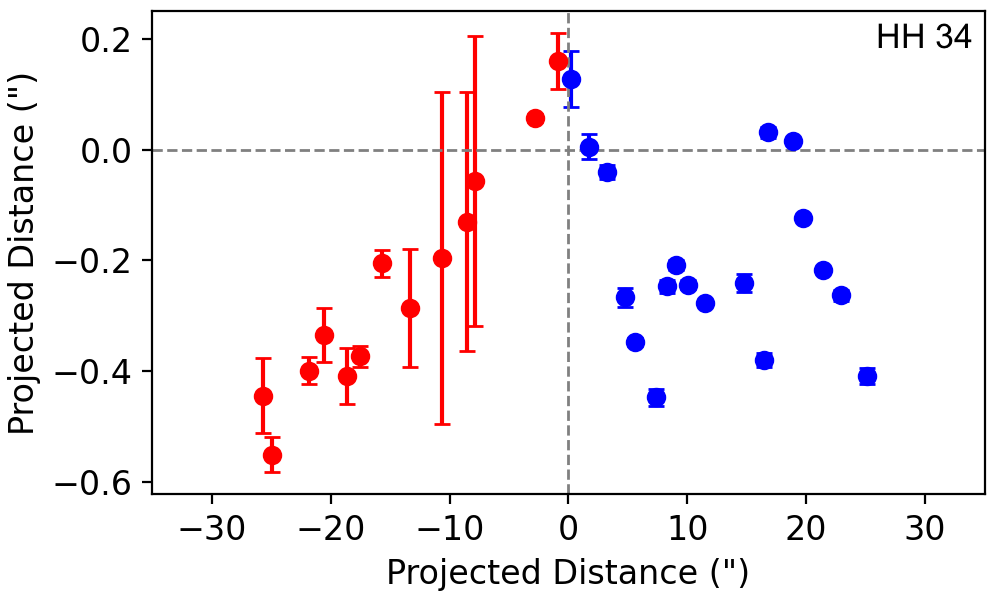}
    \\
    \includegraphics[width=0.95\columnwidth]{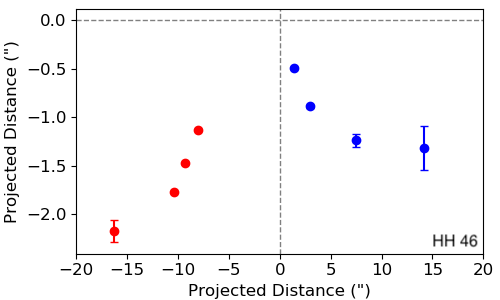}
    \includegraphics[width=0.95\columnwidth]{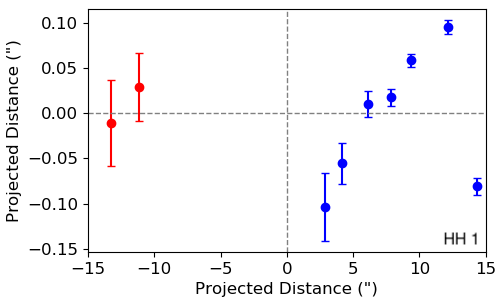}%%
    \caption{Position of the 2D photocentres of knots along the jets with respect to the jet axis and the plane perpendicular to it passing through the source (horizontal and vertical grey dashed lines, respectively). Error bars are 1$\sigma$.}
    \label{fig:jet_axis}
\end{figure*}

Since we observe both jet and counter-jet in all sources, we can investigate asymmetries between them. 
%to identify possible causes of the jet axis asymmetries (i.e. precession of the disk or orbital motion in a binary system). 
%The continuum-subtracted [\ion{Fe}{2}] 1.64~$\mu$m images were rotated such that the inner knots were aligned to the horizontal.
%The continuum-subtracted [\ion{Fe}{2}] 1.64~$\mu$m images were masked at 3$\sigma$. 
To this aim, we have mapped the jet and counter-jet trajectory, identifying the position of each jet knot photocentre by using a 2D Gaussian fit. The errors on the knot centroids were found using 1D Gaussian fits across the jet. 
%\jess{JE: The 2D fits were done in imexam which didn't give the 2D errors so this is why I used the 1D errors from fitting the widths.} 
%The relative positions of knots with respect to the central driving sources (whose position is taken from VLA or ALMA observations) have been measured for all the jets. 
We considered only the knots in the main body of the collimated jets, as they are more compact and thus their centroids are less prone to uncertainty caused by the presence of diffuse emission, unlike in extended bow shocks at the jet apex. The knot positions with respect to the jet axis (defined according to the PA given in Table 1) are plotted in Figure \ref{fig:jet_axis}. The driving source position has been taken as the origin of the jets. All jets show a systematic displacement with respect to their axis. For HH~34 and HH~111, the driving source position, taken as coincident with the VLA sources, appears shifted by about 0$\farcs$1 with respect to the jet axis. It is unlikely that this is an effect due to extinction, since we use the radio source coordinates which should not be affected by diffuse emission, therefore this apparent shift indicates a possible misplacement of the source coordinates with respect to the \textit{HST} images. 
%, and the detection of the counter-jets allows us to observe that the wiggling has a mirror symmetry with respect to the central source. 

In HH~111, HH~34 and HH~46, we observe symmetry between the corresponding blue- and red-shifted knots with respect to the plane perpendicular to the jet passing through the source. In particular, the HH~111 jet is very well traced at distances up to $\pm$ 60$\arcsec$ from the central source and both the jet and counter-jet show mirror symmetry in their trajectories.
%wiggling pattern with an amplitude that increases with distance. 
This configuration can be caused by the orbital motion of the jet source around a companion, as we will discuss in Section \ref{sec:asymmetry}. In HH~34 and HH~46, while the blue- and red-shifted knots also appear to show mirror symmetry, we cannot identify a clear undulation with a jet/counter-jet symmetry. Therefore, we cannot exclude other possible causes of the observed pattern given the large error bars in the red-shifted knots (for HH~34) and the low number of knots (for HH~46).
For HH~1, an insufficient number of red-shifted knots are observed to allow differentiation between mirror symmetry and point symmetry patterns in the trajectory. Note, however, that identifying the exact jet PA using the inner knots is tricky, and can significantly influence the interpretation of the asymmetry. We will discuss the observed asymmetries in more detail in Section \ref{sec:HH111}. 
%It should also be noted that determining the correct alignment can be quite difficult, and can heavily influence the interpretation of the asymmetry. 

%\begin{itemize}
%    \item where counter jet is seen (HH~34, HH111) look for asymmetries in jet axis position (wiggling) and PA - could be useful for disk people (emphasise this?)
%    \item "Fold" images in half and "stretch" jets using proper motions info to examine point or mirror symmetries and also to see if ejection happens simultaneously in both lobes.
%\end{itemize}

%\subsection{Emission Line Ratios}

%Fe/OI ratio – Fe image size $\approx$ 1000x1000 pixels, OI images $\approx$ 4000x4000 pixels. Region covered by Fe image using a box region in DS9 \& copied this to the OI image to see where (in WCS coordinates) they overlapped. Cropped the OI image to the same aspect ratio as the Fe images, to be rebinned to $\approx$ 1000x1000. Multiply each image by Weff (in $\AA$) for that filter, create ratio maps with imarith.

\subsection{Extinction along the jets}
%\color{red}
%\begin{itemize}
 %   \item Describe the method to measure extinction from the [FeII] line ratio
 %   \item Explain difficulties in the measurement, expecially close to the source where residual of continuum emission remains
%    \item Generally redder moving closer to the source. This is expected because of the cavity/nebulosity being thicker near the source and the disk.
%   \item Averaged extinction for knots in jet to produce 1D curve of Av along jet and only considering points where R$_{obs}$ is greater than R$_{theor}$ (=1.1 Giannini et al 2015). y error is 14\% based on WFC3 PHOTFLAM calibration errors of 10\%, x error is range over which Av was averaged.
%    \item make comparison between cont-subt and non-cont-sub profiles. We should give an estimate on how much the presence of emission lines in the continuum filters affect the final result
%\end{itemize}

%Fe ratio \& Av – Multiply each image by filter effective width (in Å) for that filter to convert flux unit from erg/cm2/s/A to line flux units erg/cm2/s . imarith in IRAF to create ratio file, then masked ratio to pixels > 3xRMS and calculated Av assuming intrinsic 1.25/1.64~$\mu$m ratio 1.1 (Giannini et al., 2015).

%\color{black}

The [\ion{Fe}{2}]1.25~$\mu$m/1.64~$\mu$m line ratio is independent on the gas physical conditions (temperature and density), since the two lines originate from the same upper level. Consequently, this ratio depends only on the atomic physics and on the reddening along the line of sight. We can therefore use our maps obtained in the two narrow band filters to estimate the extinction along the jet. This method assumes that the emission entering our narrow band filters is only due to the reddened lines. However, as already noted, close to the source there is a significant contribution from nebulosity due to continuum scattered light. Therefore, we measured the line ratio using images which were first continuum-subtracted. 
%This ratio was computed using the continuum-subtracted images, because the continuum due to the reflection nebula dominates close to the source.
The presence of fainter [\ion{Fe}{2}] lines falling within the bandwidth of the continuum filter can also contaminate the continuum-subtracted flux measurements. We evaluate this contamination in the Appendix~\ref{appendix_c} and estimate it to be 5\% at most in the dense jet sections close to the source. 

\begin{figure}[ht]
   \centering
    \includegraphics[width=0.95\columnwidth]{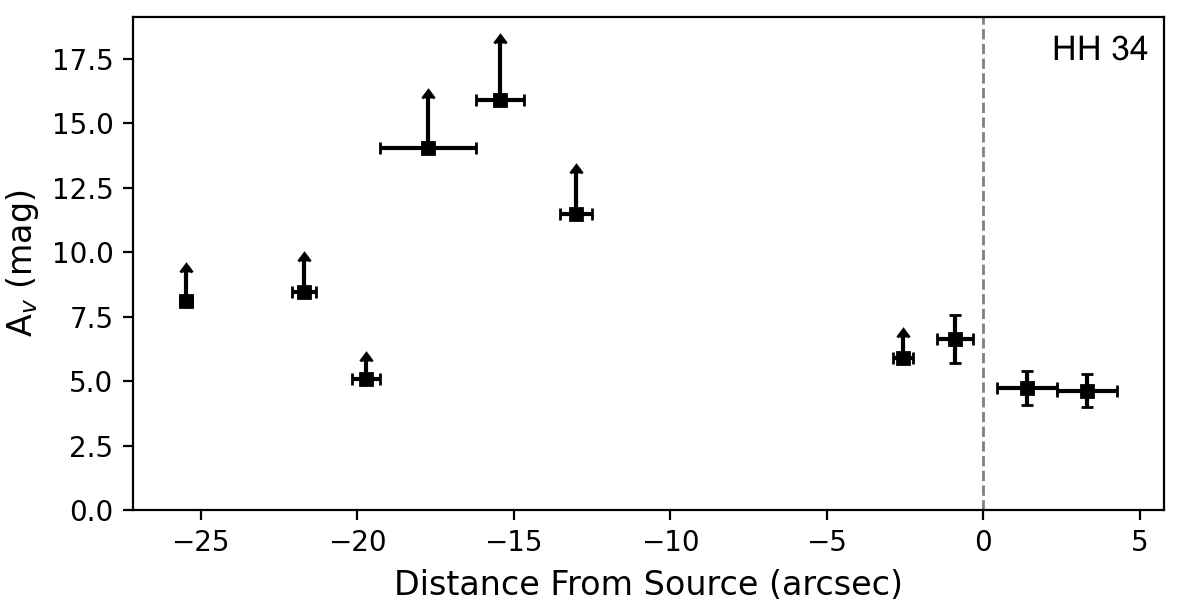}
    \includegraphics[width=0.95\columnwidth]{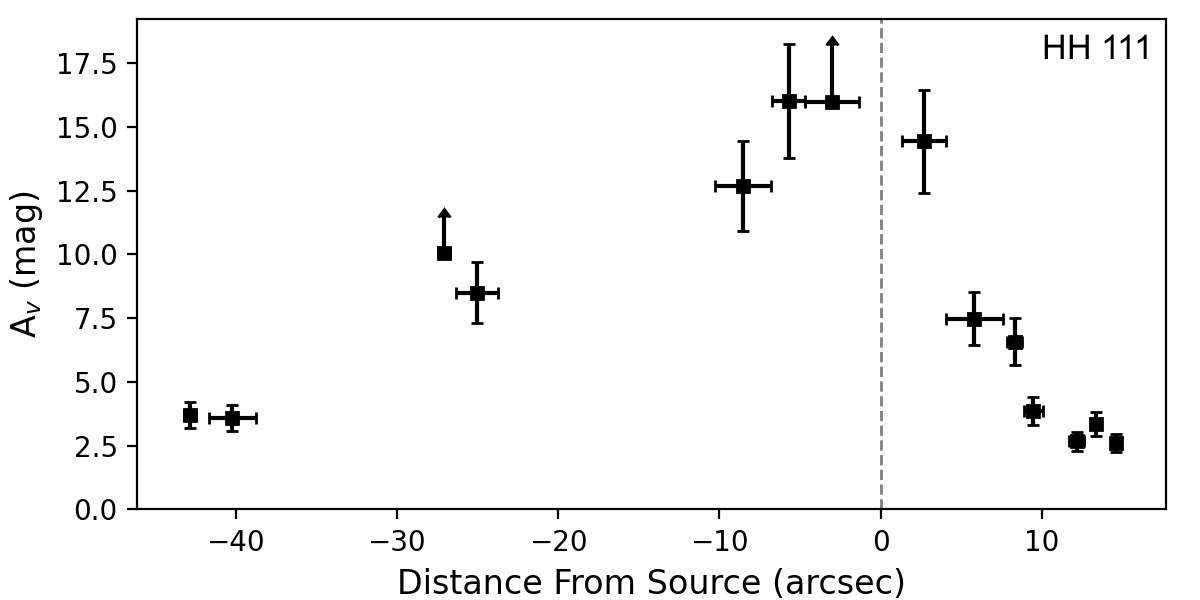}%%
    \\ \includegraphics[width=0.95\columnwidth]{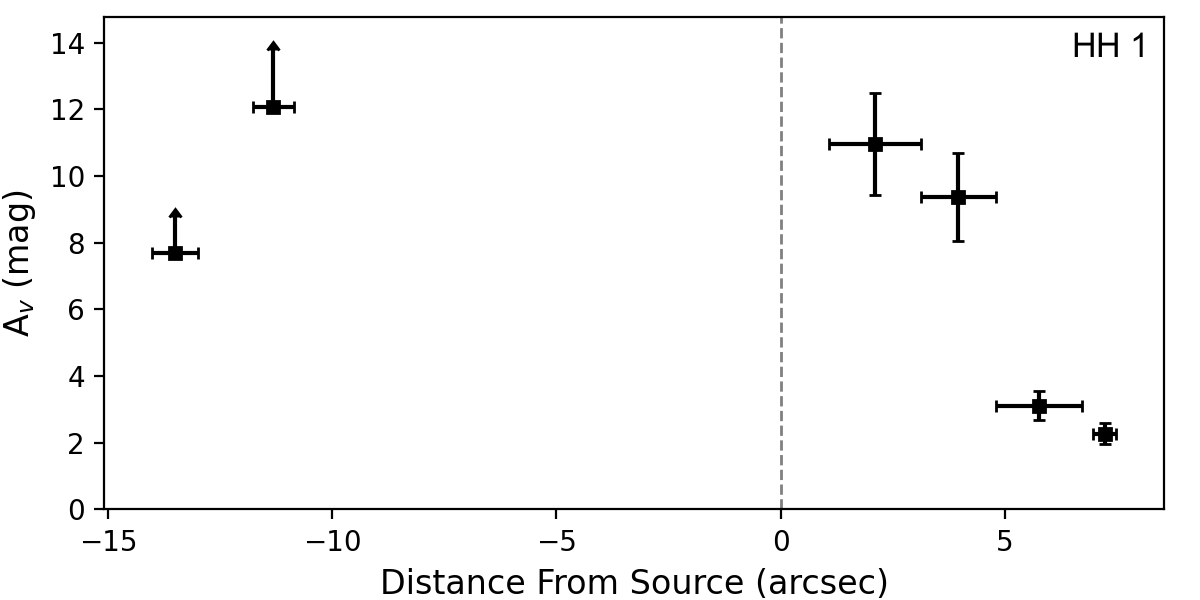}
    \includegraphics[width=0.95\columnwidth]{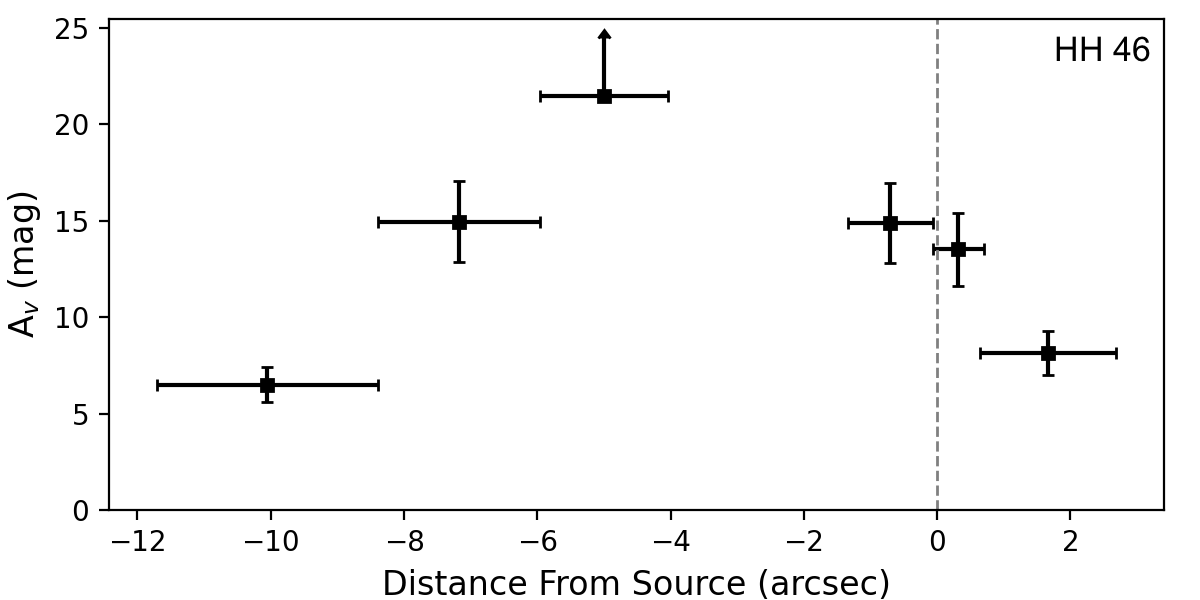}%%
    \caption{Extinction curves for all four sources. Extinction values calculated from averaged flux values of the [\ion{Fe}{2}] 1.25~$\mu$m and 1.64~$\mu$m lines, assuming a theoretical ratio value of 1.1 \citep{Giannini2015b}. Upper limits were calculated at points where the SNR of the 1.25~$\mu$m line was less than 3.}
    \label{fig:extinction}
\end{figure}

%The continuum-subtracted images were multiplied by the filter effective width (in \AA) to convert flux units from ergs cm$^{-2}$ s$^{-1}$ \AA $^{-1}$ to line flux units ergs cm$^{-2}$ s$^{-1}$. 
%The images were masked to 3$\sigma$ 
To measure the line ratio along the length of each jet, the jet images were binned across the width of the jet, producing the 1D line flux curves in Figure \ref{fig:extinction}. The extinction values were calculated by averaging the flux in each knot along the jet and assuming the empirically determined intrinsic value for the [\ion{Fe}{2}]1.25/1.64~$\mu$m ratio of 1.1 \citep{Giannini2015b}. The \citet{Cardelli1989} extinction law has been used to estimate $A_V$.
The errors in the extinction calculation were taken to be 14\% based on the {\it HST}/WFC3 \textit{PHOTFLAM} calibration errors of 10\%. 

Figure \ref{fig:extinction} and Table \ref{tab:extinction} show high visual extinction values close to the source, typically 10-15 mag, that gradually decrease towards negligible values at further distances. In the inner regions, residuals of continuum emission and noise introduced by continuum subtraction may influence the results. Red-shifted counter-jets have consistently larger values with respect to the corresponding blue-shifted jets. In HH~34 and HH~1, where the counter-jet has been detected only in the 1.64~$\mu$m line, lower limits on the A$_V$ of about 10 mag have been estimated.

\begin{table*}[ht]
\caption{Extinction (A$_{\nu}$) values along each of the four jets. The error is 14\%.}
\begin{tabular}{||c|c||c|c||c|c||c|c||}
\hline
\multicolumn{2}{||c||}{{HH 1}} & \multicolumn{2}{|c||}{{HH 34}} & \multicolumn{2}{|c||}{{HH 46}} & \multicolumn{2}{|c||}{{HH 111}} \\ \hline
z ($"$) & A$_{\nu}$ (mag) & z ($"$) & A$_{\nu}$ (mag) & z ($"$) & A$_{\nu}$ (mag) & z ($"$) & A$_{\nu}$ (mag) \\ \hline
-13.5  & 7.7  &  -25.47  & 8.1  & -10.05  &  6.5 & -42.82 & 3.69      \\
-11.3  & 12.1  &  -21.7  &  8.5 &  -7.17 &  14.5 & -40.19 & 3.57      \\
2.11  & 10.9  &  -19.71  & 14.0  & -4.99  & 21.5  & -27.07 & 10.05     \\
3.97  & 9.4  &  -17.73  &  15.9 & -0.70  & 14.9  & -25.02 & 8.49      \\
5.76  & 3.1  &  -15.42  &  11.5 &  0.32 & 13.5  & -8.51  & 12.68     \\
7.23  & 2.3  &   -12.99 & 5.9  & 1.66  &  8.2 &  -5.70  & 16.01     \\
  &   &  -2.56  &  6.6 &   &   & -3.01  & 15.99     \\
  &   &  1.41  &  4.7 &   &   & 2.69   & 14.43     \\
  &   &  3.33  &  4.6 &   &   &  5.82   & 7.48      \\ 
  & & & & & & 8.32   & 6.57      \\ 
  & & & & & & 9.47   & 3.86      \\
  & & & & & & 12.16  & 2.67      \\ 
  & & & & & & 13.36  & 3.36      \\ 
  & & & & & & 14.59  & 2.60      \\ \hline
\end{tabular}
\label{tab:extinction}
\end{table*}

\section{Discussion} \label{sec:discussion}

\subsection{Comparing Class 0/I to Class II jet widths and collimation}

\begin{figure*}
    \centering
    \includegraphics[width=0.9\textwidth]{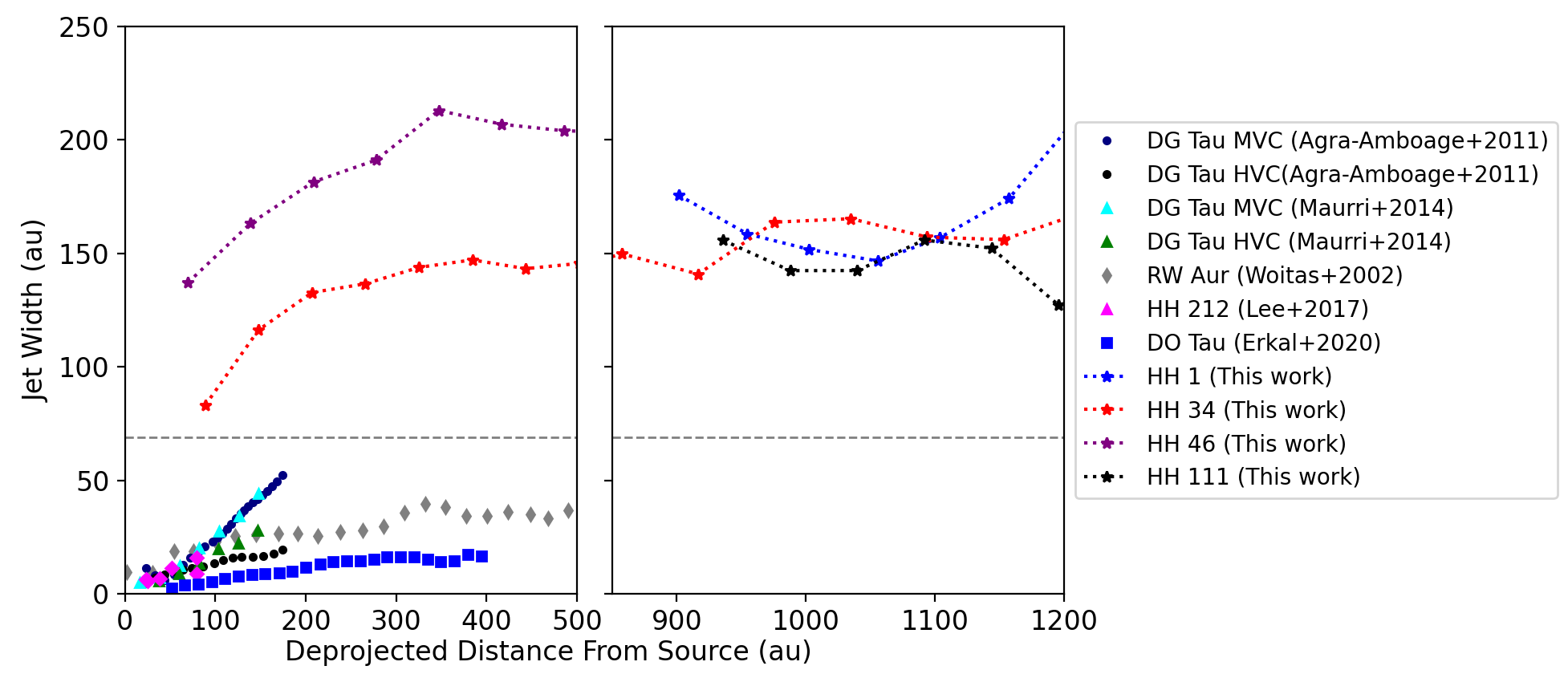}
    \caption{Comparison of Class~0/I jet widths derived in this paper with the widths measured for several T~Tauri jets in the literature. The horizontal dashed line is the instrumental FWHM (0.153$\farcs$) in our measurements.}
    \label{fig:jfwhm_compareTT}
\end{figure*}

Figure \ref{fig:jfwhm_compareTT} compares the widths of our four jets measured in [\ion{Fe}{2}] with those of T~Tauri jets in the literature which are based on high spatial resolution observations (i.e. using {\it HST} or using ground-based observations with adaptive optics). 
The HH~1 and HH~111 widths appear only in the right panel of Figure \ref{fig:jfwhm_compareTT} because we do not observe these jets close to the source. Similarly, the HH~46 jet is not shown in the right panel because it becomes wider than the y-axis range of the plot. 
%We see the jets are highly collimated with opening angles of less than 10$^{\circ}$ (see Table \ref{table:angle}). 

%Our results directly compare to those of \citet{AgraAmboage2011} for DG Tau and \citet{Erkal2021} %for DO Tau, which are based on the same  [\ion{Fe}{2}] tracer. 
%Meanwhile, other literature results plotted are based on different jet tracers. However, these still %provide useful comparisons since the majority measure the jet widths in [\ion{S}{2}] which traces the %same parts of the jet as [\ion{Fe}{2}] as we discussed in previous section. 

Our results directly compare with those of the T~Tauri jets, which are based on [\ion{Fe}{2}] or [\ion{S}{2}] lines tracing the same parts of the jet, as we discussed in previous section.

As seen in Section~\ref{section:width}, HH~34 and HH~46 have opening angles of $\sim$ 1.5 and 8.5 $^{\circ}$ (respectively) and this collimation, similar also for the other jets, is preserved up to very large distances. 
In HH~34, however, the widths measured in the first points at z $<$ 400 au suggest an initial wider opening angle of $\sim$ 7$^{\circ}$ that however would need confirmation with observations at higher spatial resolution.
%A similar behaviour of initial large opening angle followed by a slower increase in width with a much smaller opening angle of a %few degrees has been observed in several T~Tauri jets (e.g. \citet{Ray2007}). In this latter case  the transition between the two %collimation regimes occurs on much smaller scales, i.e. 20-50 au, and this change is taken as an indication that the jet initially %expands with a wide angle and it is then recollimated 

The more striking difference emerging from Figure \ref{fig:jfwhm_compareTT} is that the HH~34 and HH~46 jet widths are much wider than for the T~Tauri sources when compared on the same spatial scale. T~Tauri jets actually show a wide range of jet widths, which also depend on the velocity component that one is analysing. In DG Tau, for example, spectro-imaging observations show that the jet component at high velocity (the HVC in Figure \ref{fig:jfwhm_compareTT}) is narrower and more collimated than the component at medium velocity \citep[the MVC][]{AgraAmboage2011,Maurri2014}. In any case, the velocity integrated widths measured on HH~34 and HH~46 are a factor of 2-3 larger than those estimated for the wider DG Tau velocity component. 
From Figure \ref{fig:jfwhm_compareTT}, we also note that the widths measured on the Class 0 source HH~212 by means of SiO observations \citep{Lee2017} is in line with the values measured on T~Tauri stars, which suggests that jet collimation does not necessarily depend on evolution. 

Given our measured widths at a distance of $\sim$ 100 au, a simple linear extrapolation of the observed jet opening angle back to the disk plane would project an initial diameter of about 70~au and 125~au for HH~34 and HH~46, respectively. These represent upper limits on the jet launching regions, if one considers magneto-centrifugal mechanisms where the jet initially expands and is then recollimated at few au above the disk \citep[e.g.][]{Pelletier1992}. Estimates of jet launch radii performed through observed jet rotation give values between 0.1-4~au for the high velocity atomic component of T~Tauri jets (\citealp{Bacciotti2002}, \citealp{Coffey2004}, 2007), while for the Class I sources HH~26 and HH~72 \citet{Chrysostomou2008} measured slighting larger launch radii of 2-4 au through H$_2$ observations. Also, \textit{HST} observations of the HN Tau and UZ Tau E jets indicate an original width at the jet source $<$ 5 au \citep{Hartigan2004}.
If we assume that the HH34 and HH46 jets originate from similar launching radii at $<5$ au, then the widths measured at 100 au distance imply that the jets should initially undergo a fast expansion with an opening angle of $>$ 40 degrees.

An additional possibility is that at large distance we are not measuring the intrinsic jet diameter but rather the width of internal, unresolved, bow shocks or that the width appears larger because of the additional contribution from envelope material entrained in a turbulent mixing layer \citep[e.g.][]{Binette1999}. Observations at higher spatial resolution would be needed to explore the various possibilities. 

\subsection{Asymmetric lobes}\label{sec:asymmetry}
%\jess{\textbf{Jochen:cite additional studies of asymmetric lobes?}}
There are several studies in the literature which together show that protostellar jets are rarely well-centred on their propagation axis. In many cases, jet knots exhibit a regular wiggling pattern that cannot be explained by a change in the trajectory of the jet due to obstacles along the path. Alternatively, observed wiggling may be due to variations in the direction of the ejection at the jet origin. Such direction changes can have various causes. The possibility that the observed undulations are due to a misalignment of the jet axis with the source rotational axis is generally ruled out, because it would produce precession on timescales which are too short. It is more likely that the wiggling originates from the presence of one or more companions, in which case the wiggling pattern may be due either to the orbital motion of the driving source around the companion, or to precession of the disk plane due to tidal interactions in non co-planar binary systems \citep[e.g.][]{MasciadriRaga2002, Terquem1999}. These two scenarios can be disentangled through examination of the symmetry pattern of the trajectory presented by the jet and counter-jet with respect to the central source: a mirror-symmetry in the case of orbital motion, and a point-symmetry in the case of precession. 

The jet/counter-jet symmetry has been studied in many sources, both Class II (T~Tauri) and Class 0/I. Mirror-symmetric jets seem more common than point-symmetric jets \citep[e.g.][]{NoriegaCrespo2020, Estalella2012} although well-known examples of precessing jets have been observed, e.g. the Class 0 outflows L1157 \citep{Gueth1996, Takami2011} and Cep E \citep{Eisloffel1996}.
In our study, we find that three out of four targets (i.e. HH~34, HH~111 and HH~46) show mirror-symmetry in the red- and blue-shifted knot positions. However, as noted in Section \ref{sec:jaxis}, we find some evidence of a mirror-symmetry pattern caused by orbital motion of the jet source only for HH~111, due to the difficulty in identifying a clear wiggling pattern in HH~34 and HH~46. In the HH~1 jet (see Figure \ref{fig:jet_axis}), the detected red-shifted knots are too few and faint to ascertain the type of symmetry. 

The HH~46 blue-shifted outflow has a very pronounced large scale helicoidal pattern. \citet{MasciadriRaga2002} interpreted this pattern as due to the orbital motion of the jet source around the companion found by \citet{Reipurth2000b} at $\lambda \sim 2\mu$m with a separation of approximately 120 au. Both \citet{MasciadriRaga2002} and \citet{Reipurth2000b} however conclude that the complex morphology of the outflow cannot be reproduced by this simple interpretation and could require the presence of a triple system. 

For HH~34, the symmetry between the jet and counter-jet was discussed in \citet{Raga2011}, where the counter-jet at distances $>$ 5$\arcsec$ was for the first time revealed by means of Spitzer observations. They observed an offset in the positions of corresponding jet/counter-jet knots, and interpreted it as due to a velocity difference and a time delay in ejection between the corresponding knot pairs. 

\begin{figure}
   \centering
    \includegraphics[width=0.95\columnwidth]{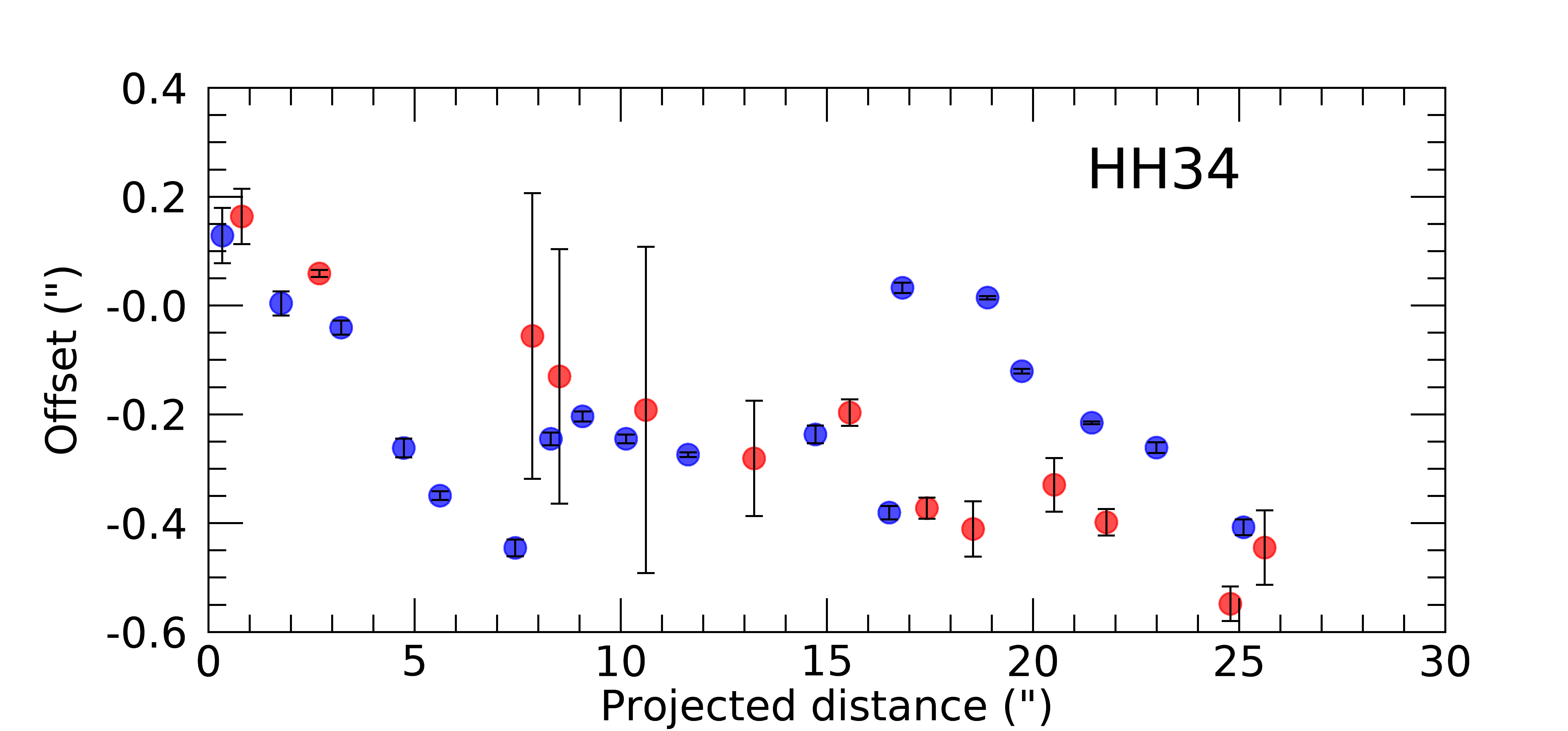}
    \caption{Offsets of the knots in the HH~34 jet (blue) and counter-jet (red) with respect to the jet axis originating from the VLA driving source. The counter-jet knot positions have been folded to be superimposed on the jet distance scale.} 
    \label{fig:HH34_folded}
\end{figure}

We now observe the HH~34 counter-jet at a higher spatial resolution and closer to the source than was previously possible. In order to better compare the knot displacement in the two lobes, we show in figure \ref{fig:HH34_folded} the offsets as a function of distance, where the positions of the red-shifted knots have been folded over so as to be plotted on the same x-scale as the blue-shifted knots. Note that, for both the blue and red data-points, there is a sharp increase of the offsets with respect to the jet axis for distances $ < \sim$ 10$\arcsec$. However, this increase in offsets occurs much more rapidly in the blue-shifted jet, followed by an abrupt change in direction at about 7$\arcsec$ after which the jet and counter-jet axes seem to align. It is not clear whether such a sharp kink of the jet axis exists in the counter-jet, because the corresponding region is extincted by the large nebulosity and so the jet trajectory cannot be well traced in this region. 
Although the knots do not follow a straight trajectory, it is difficult to identify any clear undulation with a jet/counter-jet symmetry. Jet/counter-jet knot positions are also slightly shifted with respect to each other, but we cannot find evidence of any regular pattern in these shifts. 

\subsubsection{Modelling the orbital motion of HH~111}\label{sec:HH111}

The HH~111 jet shows a clearer symmetry between jet and counter-jet, with a defined mirror-symmetric pattern. Such a symmetric pattern was noticed by \citet{NoriegaCrespo2011}, who studied the positional distribution of the jet knots in both lobes observed by Spitzer. Thanks to our higher spatial resolution, we can interpret the observed pattern as orbital motion of a binary system. 

We model the spiral pattern seen in HH~111, adopting the formulation given in \citet{Anglada2007} and \citet{MasciadriRaga2002} for a ballistic jet of a star in circular orbit, assuming a constant jet velocity. 
The parameters that enter into the model are directly linked to observable quantities. In particular, the orbital radius of the jet source around the binary centre of mass, $r_o$, is given by: 

 \begin{equation}
  r_o = \frac{\lambda\,\rm{tan}\,\alpha}{2\pi}\,D
 \end{equation}
 
 where $\lambda$ is the angular distance in the plane of the sky between the positions of two maximum elongations, $\alpha$ is the half opening angle of the jet and $D$ is the distance. 
 In addition, if $v_j$ and $v_o$ are the velocities perpendicular and parallel to the orbital plane, their ratio is given as:
 
  \begin{equation}
  k = v_j/v_o = {\rm tan}\,\alpha\,{\rm cos}\,\phi = v_t/v_o
 \end{equation}
 
where $\phi$ is the inclination angle of the jet with respect to the plane of the sky, and $v_t$ is the jet tangential velocity. 

Once $r_o$ and $v_o$ are measured, the orbital period $\tau_o$ can be also given by:
  \begin{equation}
  \tau_o = 2\pi\,r_o/v_o
 \end{equation}

In addition, if $m_1$ is the mass of the jet source and $m_2$ the mass of the companion, we have :
   \begin{equation}
 m = m_1 + m_2 = \mu^{-3}\,{r_o}^3\,{\tau_o}^{-2}
 \label{eq:m}
 \end{equation}
 where $\mu$ is $m_2/m$.
 Finally, the binary separation $a$ is given by $r_o/\mu$.

For our model of the the HH~111 jet, we assume a distance of 400~pc and $v_t$ = 320 \kms, i.e. the jet velocity at origin, estimated by extrapolating the tangential velocities measured along the jet in Section \ref{section:time_var}. An opening angle and $\lambda$ value of 2$^{\circ}$ and 20$"$, respectively, are given as initial guesses for the model. Finally, we allow a slight change in the position of the driving source and the jet inclination in order to adjust the alignment with the jet axis. 

\begin{figure}
    \centering
    \includegraphics[width=1\columnwidth]{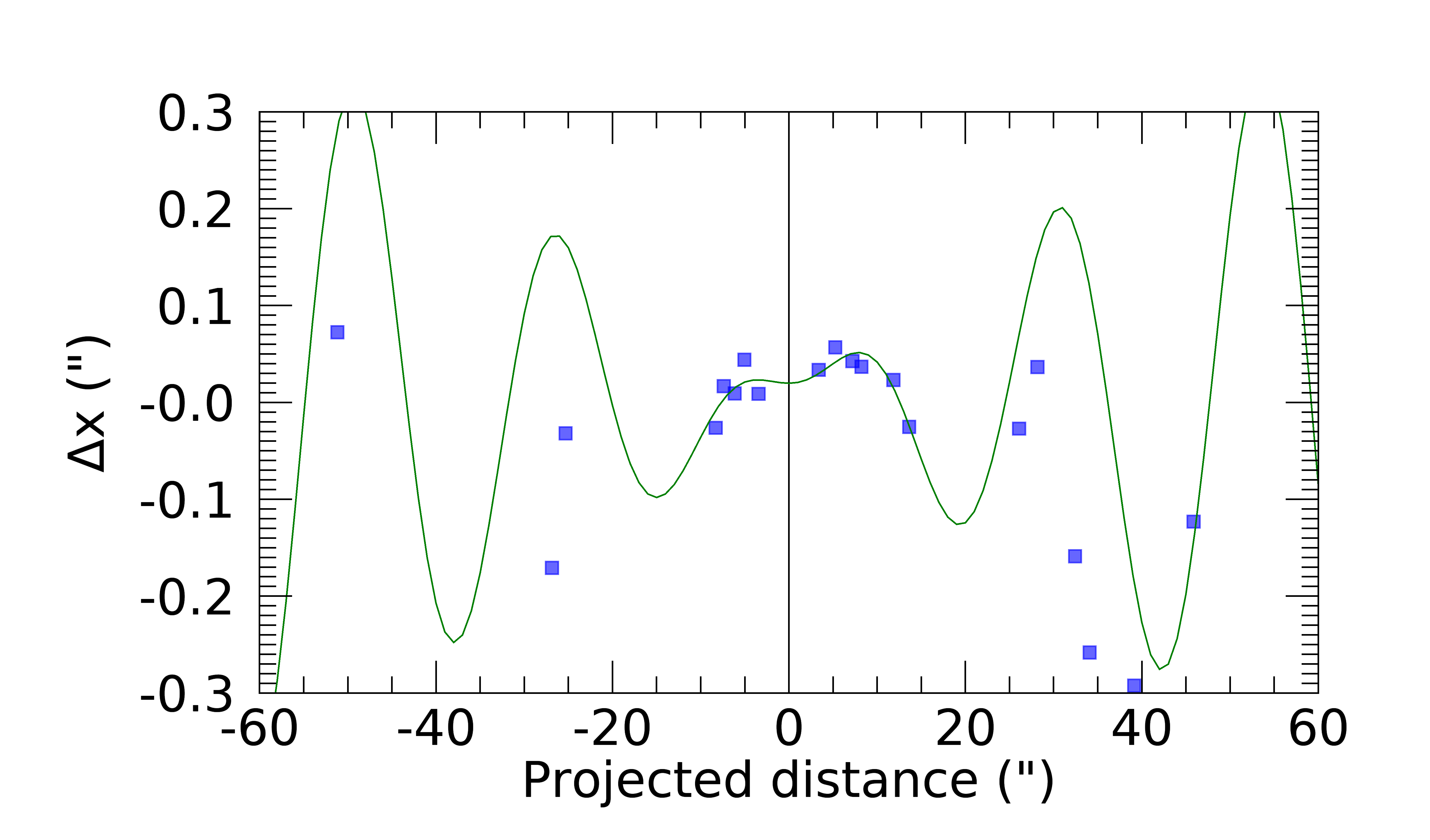}
    \caption{A model fit to the observed photocentre offset positions of the HH~111 inner knots as a function of distance from the exciting source. Blue boxes indicate the measured offsets. The solid green line is the best fit of the internal knots wiggle produced by an orbital motion of the jet source around a companion.}
    \label{fig:orbital_model}
\end{figure}

We first tried to fit the observed offsets measured along the $\pm$ 60$\arcsec$ length of the jet with the binary orbital model described above. However, we could not find a good solution that reproduced the observed wiggling for all the knots. We then fitted only the internal knots ($\pm$ 20$\arcsec$), and the result is shown in Figure \ref{fig:orbital_model} where the best fit is superimposed on the observed offsets. We note that the more external knots show the same amplitude and period predicted by the fitted curve, but they are out of phase with respect to expectations. This behaviour is consistent with the decrease in velocity seen in Figure \ref{fig:velocities_all}.
%If we assume the same acceleration pattern also for the internal knots, then the tangential velocity at the source position should be around 300 \kms. 

\begin{table}[ht]
\centering
\caption{Parameters of the orbital motion model for the inner knots.}
\begin{tabular}{|lcc|}
\hline
\hline
\multicolumn{2}{c}{Assumed parameters}\\
\hline
 $v_{t}$          & 320 \kms	&      \\
 M$_{tot}$        & 1.0-2.0 M$_\odot$ & Lee et al. (2020)\\
 $D$              & 400 pc        & Lee et al. 2016 \\
 \hline
 \multicolumn{2}{c}{Derived parameters}\\
 \hline
 $m_1$     & 0.8-1.4 M$_\odot$ & \\
 $m_2$     & 0.4-0.6 M$_\odot$ & \\
 Separation $a$   &   28-33 au & \\
 Period $\tau_o$  & 140 yr  & \\
 Orbital velocity $v_o$  & 2.0 \kms & \\ \hline
\end{tabular}
\label{tab:Param}
\end{table} 

The model fit gives as output $v_o$, $r_o$ and $\tau_o$.
ALMA observations give an estimate for the mass of the VLA~1 source of 1.5$\pm$0.5 M$_\odot$. Assuming this value as the total mass of the binary, we derive, from equation~(\ref{eq:m}) the masses of the individual sources and their separation (a=$r_o$/$\mu$). 

%Assuming that the phase difference we observe between our model and the outer knot position is due to a speed variation, we can calculate the speed difference between the outer and the inner knots that would reconcile the model with the observations. 
%If we assume an average velocity variation of XX/arcsec along the jet, then the difference in velocity between the internal and external knots considered in our analysis should be about XX \kms 

Table \ref{tab:Param} summarises all the assumed and derived parameters that we find by fitting the binary orbital model to the offsets of the inner knots for HH~111. Allowing for the total mass to lie in the range 1 - 2 M$_\odot$, the mass of the primary is in the range 0.8 - 1.4 M$_\odot$ and the mass of the secondary is in the range 0.4 - 0.6 M$_\odot$, while their separation ranges between 28 and 33 au. Such a separation is below the resolution achieved by the most recent ALMA observations \citep{Lee2020}. These ALMA observations detected a pair of symmetric spiral structures in the massive disk of VLA1.  The authors discuss the origin of these structures as due to either gravitational instabilities or the presence of a companion. In this latter scenario, the separation of the companion with respect to the primary source is estimated to be $\sim$ 40~au. Our results would thus be consistent with this proposed hypothesis. 

\citet{NoriegaCrespo2011} modelled, with a binary orbital motion, the jet displacements observed in a Spitzer 4.5 $\mu$m image. They found that the jet wiggling, observed on scales between 15 and 300$\arcsec$, was consistent with an orbital motion of a binary formed by two $\sim$ 1 M$_\odot$ stars with a separation of 186~au and an orbital period of 1800~years. Such a binary separation and period would have been produced in the inner jet a much larger $\lambda$ step, inconsistent with our \textit{HST} observations. In addition, such a companion would have been detected by the recent \citet{Lee2020} ALMA observations. 

%\begin{figure*}[ht]
%    \centering
%    \includegraphics[width=1\textwidth]{HH111_offsets_labels.jpg}
%    \caption{Top: image of the HH111 jet rotated by 47.4 degrees. Blue and red dots indicate the photocenters of the different knots used for our analysis. Bottom: Displacement of knots position with respect to the jet axis. The plot (0,0) origin indicates the position of the VLA1 source (Lee et al. 2011).}
%    \label{fig:HH111_offset_labels}
%\end{figure*}

%%%%%%%%%%%%%%%%%%%%%%%%%%%%%%%%%%%%%%%%%%%%%%%%%%%%%%%%%%%%%%%%%%%%%%%%%%%%%%%%
\section{Conclusions} \label{sec:conclusion}

We have presented {\it HST}/WFC3 images of the protostellar outflows HH~1/2, HH~34, HH~46/47 and HH~111 acquired in near-IR narrow band filters centred on the [\ion{Fe}{2}] 1.64~$\mu$m and 1.25~$\mu$m lines. The acquisition of images in adjacent filters has allowed us to construct continuum-subtracted emission line images which trace the jets as close to their origin as $\sim$ 50 au for the less obscured objects (i.e. HH~34 and HH~46). In all sources, we clearly detect several knots of the counter-jet in the [\ion{Fe}{2}] 1.64~$\mu$m line, which were barely visible or invisible at shorter wavelengths. In particular, the counter-jets of HH~1 and HH~34 are not detected even at 1.25~$\mu$m, testifying to a large envelope and cloud obscuration. 

The infrared images were used to measure key properties of these jets including proper motions, jet widths, wiggling patterns, and extinction. The main results can be summarised as follow:

\begin{itemize}
    \item By comparing our [\ion{Fe}{2}] 1.64~$\mu$m images with archival [\ion{S}{2}] {\it HST} images taken more than 10 years before, we have revised previous measurements of the jets' tangential velocities, finding values of the order of a few hundred km~s$^{-1}$ for each jet, consistent with previous measurements to within 20-30~km~s$^{-1}$. 
    \item The continuum subtracted [\ion{Fe}{2}] 1.64~$\mu$m images have been used to determine with high accuracy the jet width from large distances down to a few tens of au close to the star. In particular, we find that the HH~46 has a wide opening angle of $\approx$ 8.5$^{\circ}$ while the HH~34 jet has an initial wide opening angle of about 7$^{\circ}$, while after $\sim$ 400 au it presents a higher collimation ($\sim$ 1.5$^{\circ}$) which is preserved up to large distances. Widths close to the source have been found to be wider than more evolved Class~II sources reported in the literature by at least a factor of two. This finding suggests that either these jets are launched from larger regions in the disk or that the jets appear wider due to a contribution from the envelope material entrained in a turbulent mixing layer.
    \item We have analysed the jet and counter-jet trajectory through measurements of knot positions with respect to the jet axis. We observe symmetry between the red- and blue-shifted knots of three of our four targets (i.e. HH~111, HH~34 and HH~46), however a clear wiggling pattern is only observed in HH~111. The analysis of the knot position asymmetries in the HH~111 inner region suggests that these are due to the presence of a low mass stellar companion located at about 20-30 au from the primary source. While the binary parameters calculated for HH~111 differ significantly from low resolution Spitzer observations, our observations do agree with the Spitzer data that a binary affects the position of the observed knots. This hypothesis is further supported by recent ALMA observations of the VLA~1 disk which revealed spiral structures possibly driven by the dynamics within a binary system.
    %\item We have analysed the jet and counter-jet trajectory through measurements of knot positions with respect to the jet axis. All the jets show knot undulation, and we observe a mirror-symmetry between knot positions in the two lobes in most sources implying that each source exists in a multiple system. The analysis of the knot position asymmetries in the HH~111 inner region suggests that these are due to the presence of a low mass stellar companion located at about 20-30 au from the primary source. While the binary parameters calculated for HH~111 differ significantly from low resolution Spitzer observations, our observations do agree with the Spitzer data that a binary affects the position of the observed knots. This hypothesis is further supported by recent ALMA observations of the VLA~1 disk which revealed spiral structures possibly driven by the dynamics within a binary system.
    \item We have calculated the extinction along the jets using the [\ion{Fe}{2}] 1.25/1.64~$\mu$m ratio. We find visual extinction values of 15-20 mag near the source which gradually decreases moving downstream along the jet. We determine that the contribution of weaker emission lines in our {\it HST} narrow band filters introduce an uncertainty of 5$\%$ at most in our calculation of the [\ion{Fe}{2}] 1.25/1.64~$\mu$m ratio. 
\end{itemize}

This work highlights the importance of high spatial resolution IR observations in understanding of the jet origin in Class~0/I sources. {\it JWST} has the ability to provide observations of these jets in the mid-IR with the same resolution that {\it HST} achieves in the optical/near-IR. In particular, with JWST it will be possible to peer even deeper in the source natal envelope through imaging of low excitation [\ion{Fe}{2}] lines such as the 5.3 and 26~$\mu$m transitions. The images presented here will represent important complementary information on the [\ion{Fe}{2}] emission at higher excitation and, furthermore, will provide information on the extinction along the jet crucial for a correct quantitative interpretation of the mid-IR emission.
%Thus, the near-IR imaging study presented here provides a valuable reference for {\it JWST}, and additionally providing information about extinction needed to correct mid-IR observations.  

\section*{Acknowledgements}
This work has been supported by PRIN-INAF MAINSTREAM 2017 ”Protoplanetary disks seen through the eyes of new-generation instruments” and by PRIN-INAF 2019 ”Spectroscopically tracing the disk dispersal evolution (STRADE)”. J. Erkal acknowledges funding from the UCD Physics Scholarship in Research and Teaching.

%%%%%%%%%%%%%%%%%%%%%%%%%%%%%%%%%%%%%%%%%%%%%%%%%%%%%%%%%%%%%%%%%%%%%%%%%%%%%%%%
\bibliography{erkalj}{}
%\nocite{*}
\bibliographystyle{aasjournal}

%%%%%%%%%%%%%%%%%%%%%%%%%%%%%%%%%%%%%%%%%%%%%%%%%%%%%%%%%%%%%%%%%%%%%%%%%%%%%%%%

\appendix 
\section{[\ion{Fe}{2}] 1.25$\mu$m and [\ion{O}{1}] 6300\AA ~ images of the jets}
\label{appendix_a}

\begin{figure}[ht]
    \centering
    \includegraphics[width=0.9\textwidth]{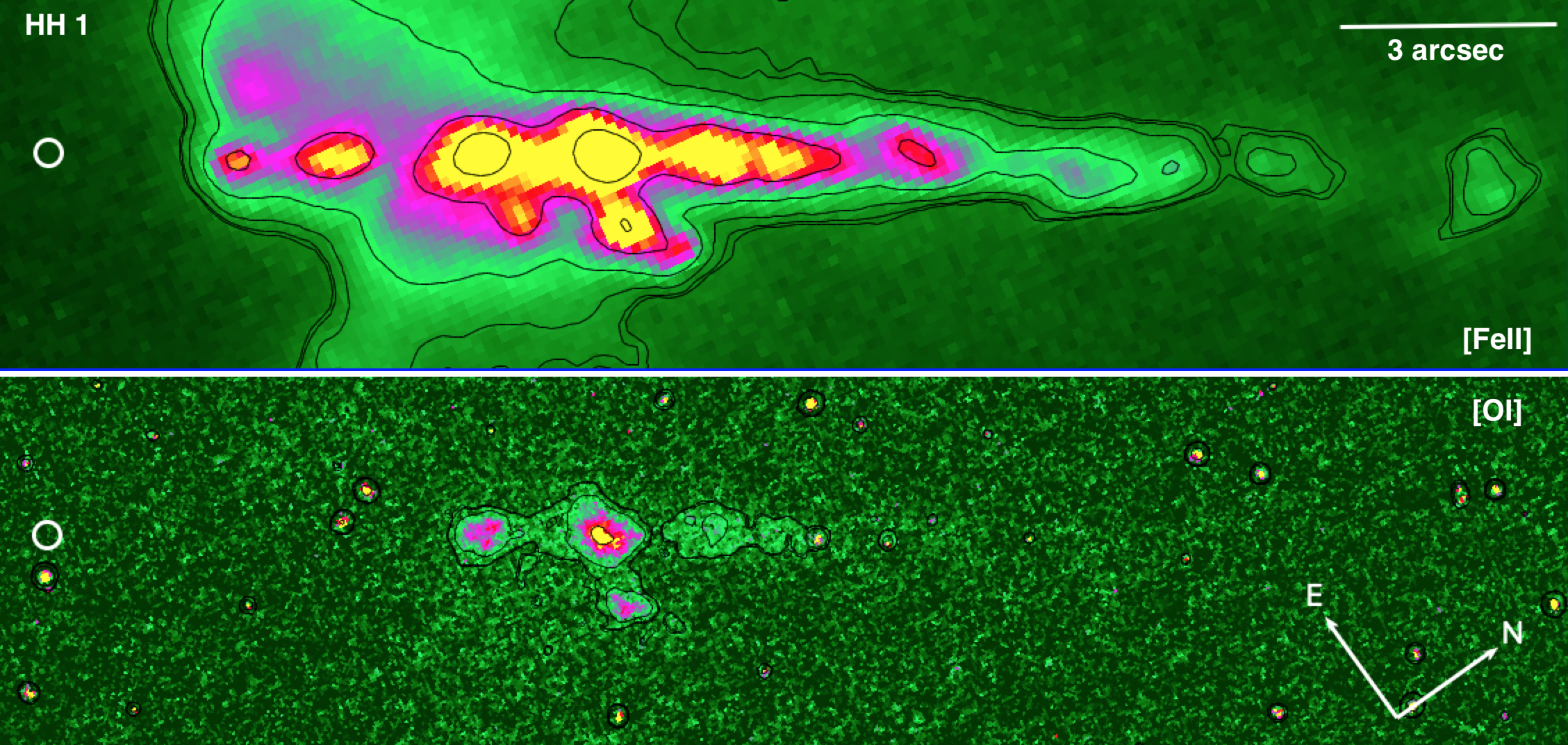}
    \caption{The HH~1 jet imaged in [\ion{Fe}{2}] 1.25$\mu$m (top) and [\ion{O}{1}] 6300\AA emission (bottom). These images are not continuum-subtracted. [\ion{Fe}{2}] contours are 0.1, 0.11, 0.13, 0.22, 0.57 and 2 $\times$10$^{-18}$ \textbf{erg\,s$^{-1}$\,cm$^{-2}$\,\AA$^{-1}$\,pixel$^{-1}$.} [\ion{O}{1}] contours are 0.5, 0.53, 0.85 and 4 $\times$10$^{-19}$ \textbf{erg\,s$^{-1}$\,cm$^{-2}$\,\AA$^{-1}$\,pixel$^{-1}$.}}
    \label{fig:hh1_fe_oi}
\end{figure}

\begin{figure}[ht]
    \centering
    \includegraphics[width=0.9\textwidth]{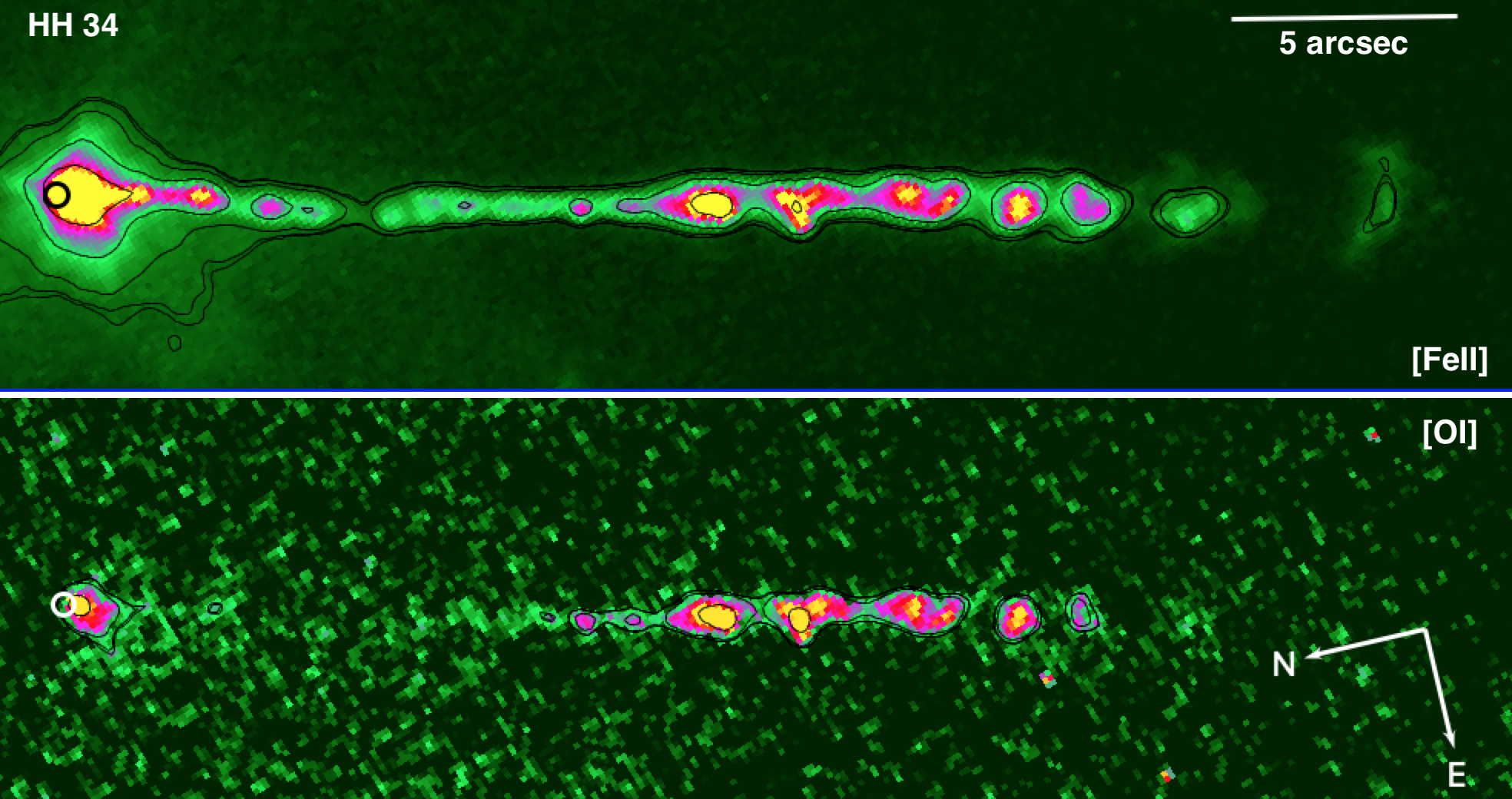}
    \caption{Same as Figure \ref{fig:hh1_fe_oi}. [\ion{Fe}{2}] contours are 0.9, 0.95, 1.2, 2.5 and 10 $\times$10$^{-19}$ \textbf{erg\,s$^{-1}$\,cm$^{-2}$\,\AA$^{-1}$\,pixel$^{-1}$.} [\ion{O}{1}] contours are 2, 2.05, 2.5 and 7 $\times$10$^{-20}$ \textbf{erg\,s$^{-1}$\,cm$^{-2}$\,\AA$^{-1}$\,pixel$^{-1}$.}}
    \label{fig:hh34_fe_oi}
\end{figure}

\begin{figure}[ht]
    \centering
    \includegraphics[width=0.9\textwidth]{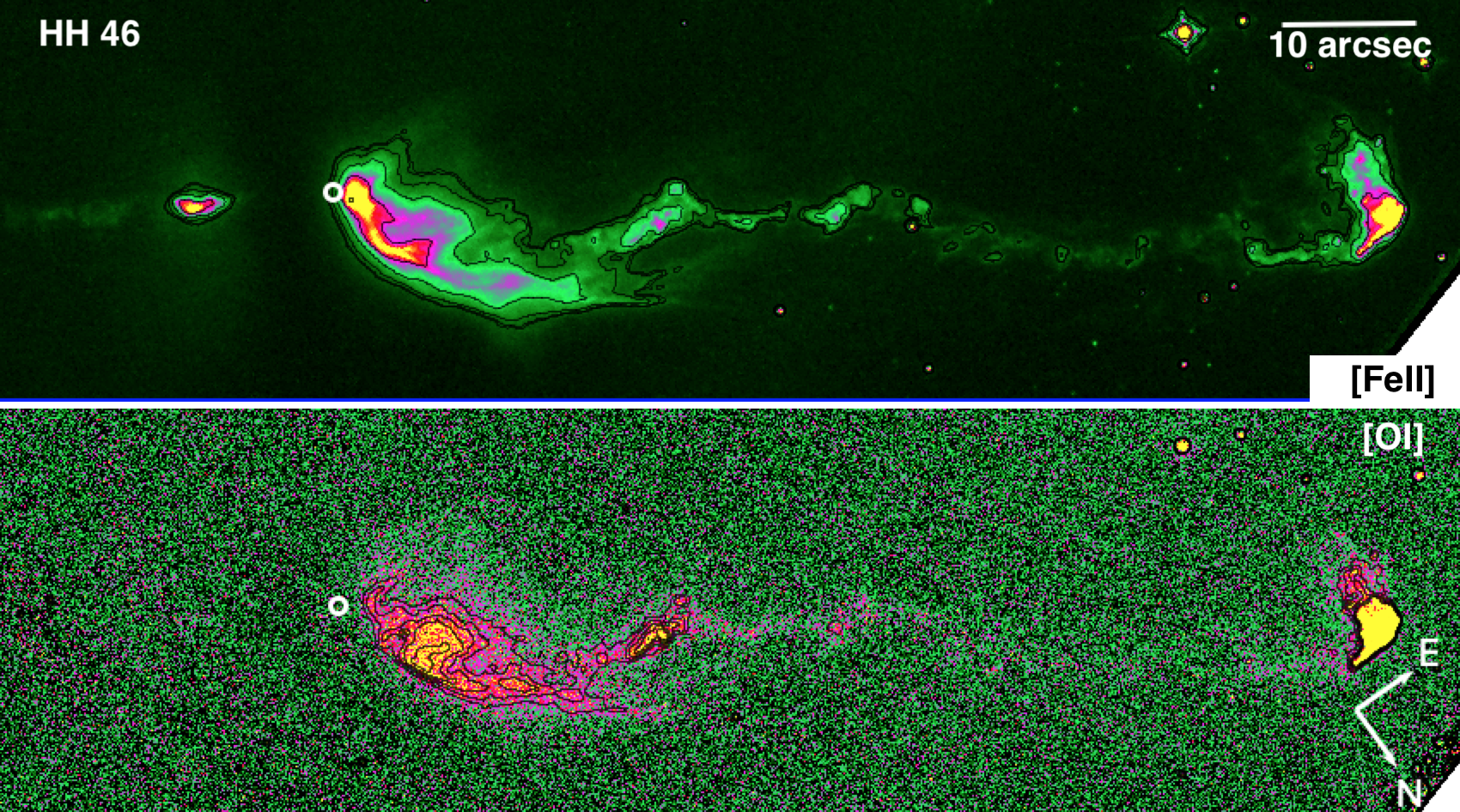}
    \caption{Same as Figure \ref{fig:hh1_fe_oi}.[\ion{Fe}{2}] contours are 0.08, 0.09, 0.1, 0.2, 0.56 and 2 $\times$10$^{-18}$ \textbf{erg\,s$^{-1}$\,cm$^{-2}$\,\AA$^{-1}$\,pixel$^{-1}$.} [\ion{O}{1}] contours are 4, 5.2, 6.5, 7.7 and 9 $\times$10$^{-20}$ \textbf{erg\,s$^{-1}$\,cm$^{-2}$\,\AA$^{-1}$\,pixel$^{-1}$.}}
    \label{fig:hh46_fe_oi}
\end{figure}

\begin{figure}[ht]
    \centering
    \includegraphics[width=0.9\textwidth]{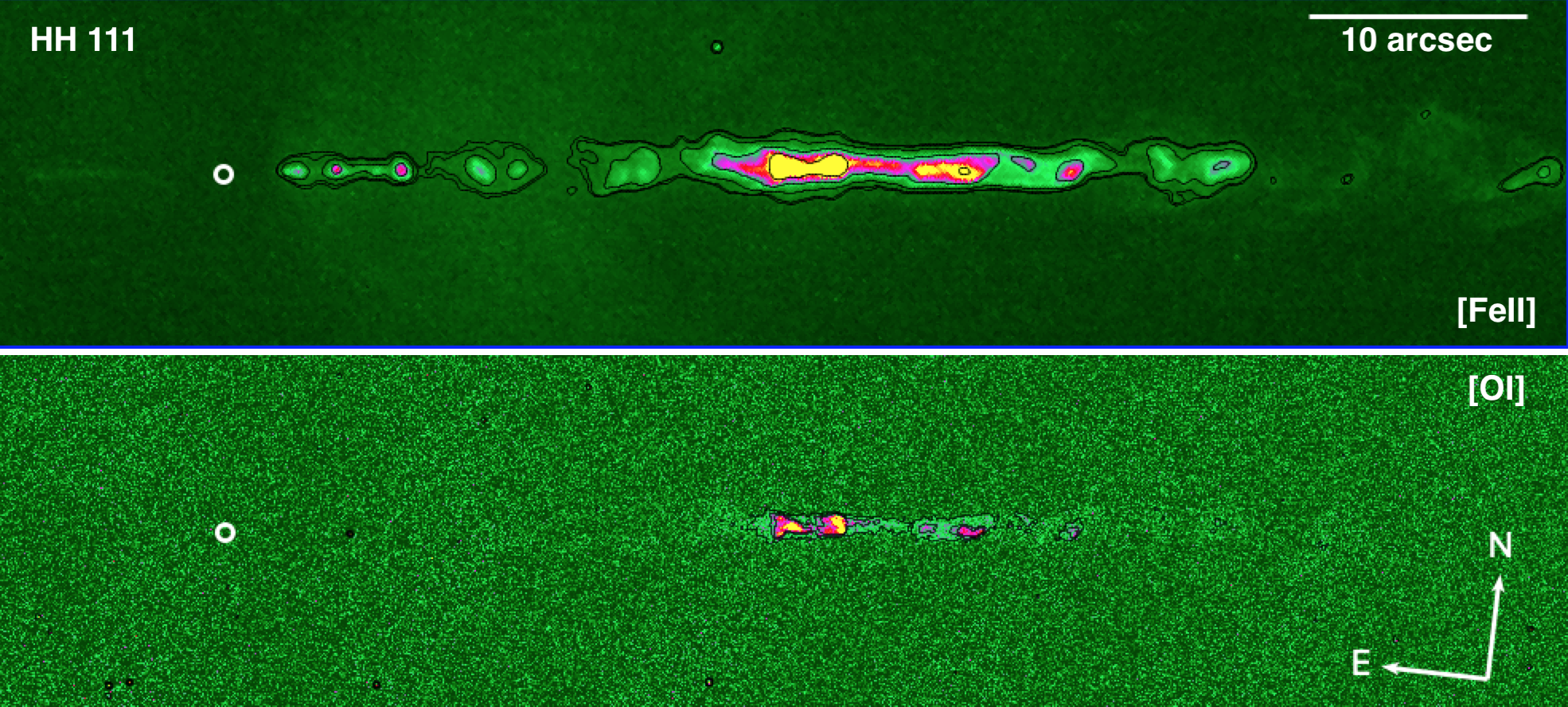}
    \caption{Same as Figure \ref{fig:hh1_fe_oi}. [\ion{Fe}{2}] contours are 0.6, 0.65, 0.9, 2.2 and 10 $\times$10$^{-19}$ \textbf{erg\,s$^{-1}$\,cm$^{-2}$\,\AA$^{-1}$\,pixel$^{-1}$.} [\ion{O}{1}] contours are 0.3, 0.65 and 1 $\times$10$^{-19}$ \textbf{erg\,s$^{-1}$\,cm$^{-2}$\,\AA$^{-1}$\,pixel$^{-1}$.}}
    \label{fig:hh111_fe_oi}
\end{figure}

\clearpage
\section{[\ion{O}{1}] 6300 \AA/[\ion{Fe}{2}] 1.64 $\mu$m line ratio maps}
\label{appendix_OIFe}

\begin{figure}[ht]
    \centering
    \includegraphics[height=0.85\textheight]{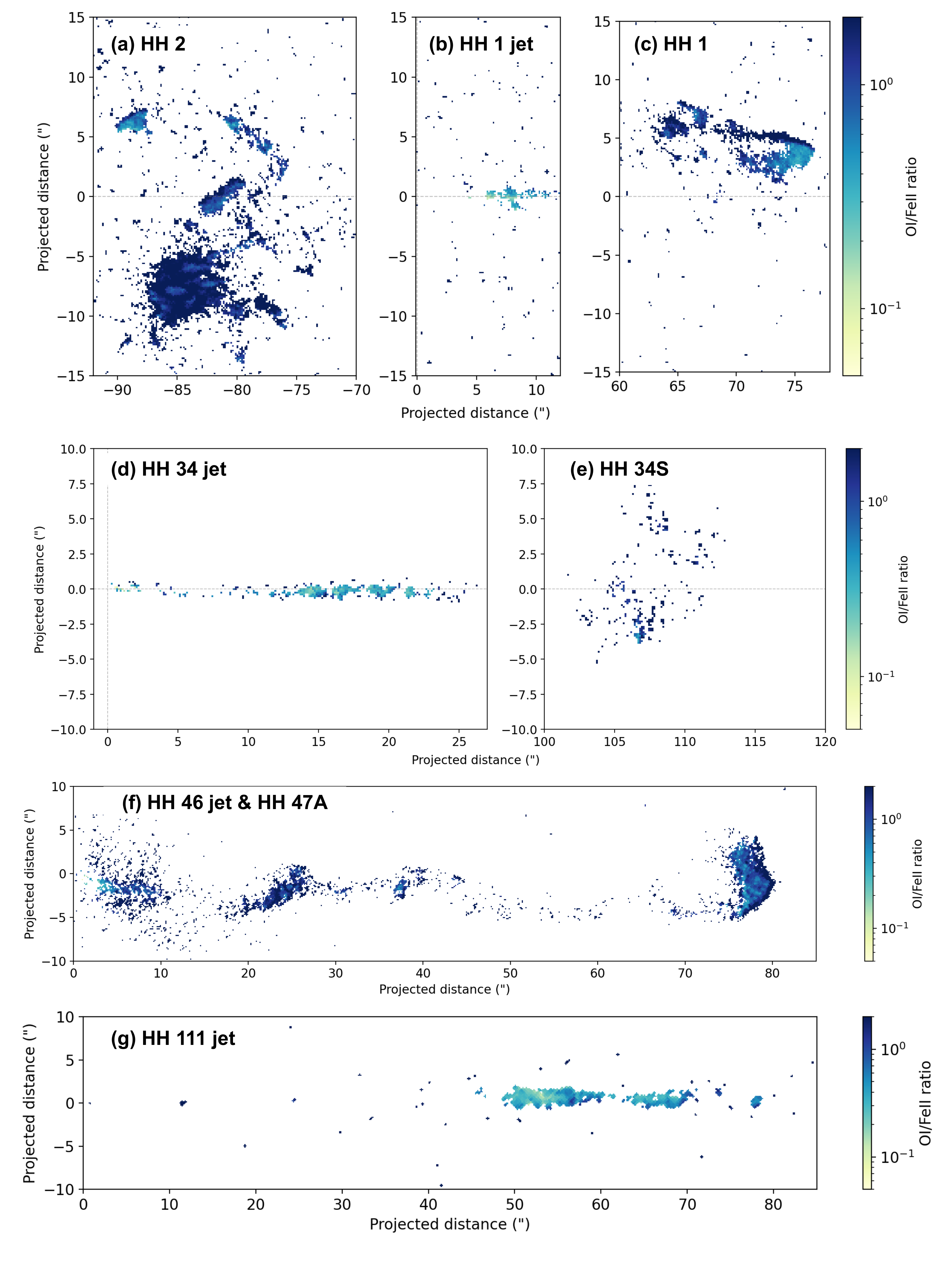}
    \caption{\textbf{[\ion{O}{1}]~6300 \AA/[\ion{Fe}{2}]~1.64 $\mu$m line ratio maps for various parts of each of our sources. Panels (a)-(c) show the [\ion{O}{1}]/[\ion{Fe}{2}] ratio for the HH~2 bowshock, HH~1 inner jet region and the HH~1 bowshock, respectively. Panels (d) and (e) show the [\ion{O}{1}]/[\ion{Fe}{2}] ratio for the HH~34 inner jet region and the HH~34S bowshock, respectively. Panel (f) shows the ratio for the HH~46 jet and bowshock HH~47A and panel (g) is the inner region of the HH~111 jet. All figures are masked above where the [\ion{O}{1}] emission is above 2 $\times$ RMS, except for HH~46 (panel (f)) where it is masked above 1 $\times$ RMS. }}
    \label{fig:oi_fe_ratios}
\end{figure}

\textbf{Figure \ref{fig:oi_fe_ratios} shows the images of the [\ion{O}{1}] 6300 \AA/[\ion{Fe}{2}] 1.64 $\mu$m line ratio in the four observed objects.
The images were obtained using the non-continuum subtracted images, to avoid introducing additional noise to the [\ion{O}{1}] images. We masked the ratio images to include only the ratio in the jet and bowshocks where both [\ion{O}{1}] and [\ion{Fe}{2}] emission are above the set RMS threshold. HH~1, HH~34 and HH~111 were masked so that only emission above 2 $\times$ RMS are present. HH~46 was masked at 1 $\times$ RMS. 
The ratios appear rather uniform in the different regions. In bow shock regions, e.g. HH~1/2 and HH~46/47, the observed ratio is typically between 0.5 and 2, while along the jets it is lower, with values of $\sim$ 0.2 - 0.3. This is mainly an extinction effect: if we correct the observed values for the extinction derived from the [\ion{Fe}{2}] line ratio, we find that the [\ion{O}{1}] 6300 \AA/[\ion{Fe}{2}] 1.64 $\mu$m ratio is consistent with what is expected from a gas with a temperature between 6000-10000~K, and electron density between 10$^{3}$-10$^{4}$ cm$^{-3}$, as estimated in the literature for these jets \citep[e.g][]{Giannini2015_hh1,Nisini2005,Nisini2016}. Spatial gradients of the ratio are also observed within the individual structures (e.g. in the HH~1 and HH~47 bow shocks) and are likely due to gradients in temperature in the post-shock regions.}

\clearpage
\section{Tables of Jet Tangential Velocities}
\label{appendix_b}

\begin{table}[ht]\label{tab:hh34_pm}
\centering
\caption{Proper motions and tangential velocities for the knots in the HH~34 jet compared to velocities reported by \citet{Raga2012}. Velocities are calculated assuming d=383~pc and \citet{Raga2012} values were corrected to d=383~pc. }
\begin{tabular}{ccccc}
\hline
Knot\footnote{Knots are numbered consecutively with distance from the source} & z & $\Delta$z  & v$_{2019}$ & v$_{2012}$ \footnote{Brackets indicate the knot nomenclature in \citet{Raga2012}}\footnote{Proper motions from \citet{Raga2012} corrected to d=383pc} \\
 & (") & (") & (\kms) & (\kms) \\ \hline
1\footnote{New knot} & 1.806 & - & - & - \\
2 & 3.195 & 1.076 & 173.4 & 171 (1) \\
3 & 4.733 & 0.918 & 148.0 & 202 (2)\\
4 & 5.651 & 0.833 & 134.3 & 171 (3) \\
5 & 7.437 & 1.057 & 170.4 & 175 (4) \\
6 & 8.305 & 1.096 & 176.7 & 164 (5) \\
7 & 9.099 & 0.919 & 148.2 & 163 (6) \\
8 & 10.141 & 0.936 & 150.9 & 141 (7) \\
9 & 11.604 & 0.937 & 151.1 & 162 (8) \\
10 & 12.722 & 0.935 & 150.7 & 165 (9) \\
11 & 14.359 & 1.005 & 162.0 & 156 (10) \\
12 & 14.730 & 0.800 & 129.0 & 156 (11) \\
13 & 16.493 & 0.791 & 127.5 & \multirow{2}{*}{142 (12)} \\
14 & 16.788 & 0.815 & 131.4 &  \\
15 & 18.873 & 0.878 & 141.5 & 148 (13) \\
16 & 19.767 & 0.856 & 138.0 & 148 (14) \\
17 & 21.454 & 0.940 & 151.5 & 148 (15) \\
18 & 22.743 & 0.918 & 148.0 & \multirow{2}{*}{138 (16) } \\
19 & 23.140 & 0.802 & 129.3 &  \\
20 & 25.049 & 0.835 & 134.6 & 133 (17) \\
21 & 29.515 & 0.793 & 127.8 & 143 (18) \\ \hline
\end{tabular}
\end{table}

\begin{table}[ht]\label{tab:hh46_pm}
\centering
\caption{Proper motions and tangential velocities for the knots in the HH~46 jet compared to velocities reported by \citet{Hartigan2005}, assuming a distance of 450 pc.}
\begin{tabular}{ccccc}
\hline
Knot\footnote{Knots are numbered consecutively with distance from the source} & z & $\Delta$z & v$_{2019}$ & v$_{2005}$ \footnote{Brackets indicate the knot nomenclature in \citet{Hartigan2005}} \\ 
 & (") & (") & (\kms) & (\kms) \\\hline
1 & 19.3 & 1.165 & 220.9 & 192 (Js2) \\
2 & 22.2 & 1.154 & 218.9 & 246 (Js3) \\
3 & 24.1 & 1.145 & 217.2 & 207 (Js6) \\
4 & 25.2 & 1.229 & 233.1 & 200 (Js7) \\
5 & 30.6 & 1.145 & 217.2 & 231 (Js10) \\
6 & 37.2 & 1.196 & 226.8 & 236 (Js12) \\
7 & 39.2 & 1.346 & 255.3 & 244 (Js13) \\
8 & 43.6 & 0.860 & 163.1 & 230 (Js14) \\
9 & 46.2 & 1.192 & 226.1 & 248 (Js15) \\
10 & 49.3 & 1.598 & 303.1 & 251 (Js17) \\
11 & 54.3 & 1.598 & 303.1 & 252 (Js18) \\
12 & 60.2 & 1.734 & 328.9 & 252 (Js19) \\
13 & 68.3 & 1.391 & 263.8 & 283 (Js20) \\
14 & 74.7 & 1.554 & 294.8 & 299 (As1) \\
15 & 78.9 & 1.107 & 209.9 & 236 (As18) \\ \hline
\end{tabular}
\end{table}

\begin{table}[ht]\label{tab:hh111_pm}
\centering
\caption{Proper motions and tangential velocities for the knots in the HH~111 jet compared to velocities reported by \citet{Hartigan2001}. Velocities are calculated assuming d=400pc and \citet{Hartigan2001} values were corrected to d=400pc. }
\begin{tabular}{ccccc}
\hline
Knot\footnote{Knots named in Figure \ref{fig:inner_regions_hh111}} & \multicolumn{1}{c}{z}& $\Delta$z & v$_{2019}$ & v$_{2001}$\footnote{Brackets indicate the knot nomenclature in \citet{Hartigan2001}}\footnote{Proper motions from \citet{Hartigan2001} corrected to d=400pc} \\ 
 & (") & (") & (\kms) & (\kms) \\\hline
\multirow{2}{*}{E} & \multirow{2}{*}{26.2} & \multirow{2}{*}{2.758} & \multirow{2}{*}{256.4} & 256 (E3) \\
 &  &  &  & 284 (E2) \\
\multirow{2}{*}{F} & \multirow{2}{*}{28.3} & \multirow{2}{*}{2.239} & \multirow{2}{*}{208.1} & 259 (F2) \\
 &  &  &  & 217 (F1) \\
G & 32.2 & 2.352 & 218.7 & 249 (G1) \\
H & 34.1 & 2.095 & 194.8 & 231 (H) \\
\multirow{2}{*}{I} & \multirow{2}{*}{36.9} & \multirow{2}{*}{2.151} & \multirow{2}{*}{199.9} & 211 (I2) \\
 &  &  &  & 228 (I1) \\
J & 38.9 & 1.935 & 179.9 & 229 (J) \\
K & 43.3 & 2.621 & 243.7 & 247 (K) \\
L & 46.1 & 2.476 & 230.2 & 192 (L) \\ \hline
\end{tabular}
\end{table}

\clearpage
\section{Contamination of various [\ion{Fe}{2}] lines in the WFC3 narrow band filters}
\label{appendix_c}

The WFC3 narrow band filters used in this study cover several emission lines whose contribution can contaminate the measurement of the [\ion{Fe}{2}] 1.25, 1.64$\mu$m line flux, in particular when considering continuum-subtracted images. Hydrogen lines of the Brackett series, like the Br~11 at 1.681$\mu$m and Br~12 at 1.641$\mu$m fall into the F167N and F164N filter band widths, respectively. However, their emission in the investigated jets is negligible, as testified by IR spectroscopy of some of them (i.e. \citealp{Nisini2005}, Podio et al. 2010). More relevant is the emission of the other numerous [\ion{Fe}{2}] lines, whose relative intensity is a function of the electron density \citep{Nisini2002}. 

Table \ref{tab:lines_cont} lists the [\ion{Fe}{2}] lines that are covered by the filters and their relative intensity with respect to the 1.64$\mu$m line for temperature $T_e$ = 10\,000 K and density $n_e$ = 10$^3$ and 10$^4$ cm$^{-3}$. At low density, the contribution of these lines within each filter is a few \% and can be thus considered negligible. However, at higher density the contribution of some of the lines falling into the continuum filters is up to 20\%. Consequently, the flux measured in the continuum-subtracted images can be underestimated by up to this factor. This is the case in the inner and denser jet region, where densities as large as 5\,10$^4$ cm$^{-3}$ have been estimated (\citealp{Nisini2005}, \citealp{Podio2006}, \citealp{Nisini2016}). 
We note, however, that the contamination of weaker lines in the F130N filter is comparable to that of the F167N filter, being in both cases a similar fraction of the 1.25 and 1.64 $\mu$m emission. Consequently, when the line ratio is estimated from the ratio of the relative continuum-subtracted images, the uncertainty introduced by the contaminating lines is not larger than 5\% at most.

\begin{table*}[ht]
\centering
\caption{\label{tab:lines_cont} [\ion{Fe}{2}] lines falling within the \textit{HST} filters}
\begin{tabular}{llll}
\hline
\hline
Line & $\lambda$ & $I(n_e = 10^3 cm^{-3})$ & $I(n_e = 10^4 cm^{-3})$ \footnote{Intensities relative to the 1.257 $\mu$m line estimated assuming $T_e$ = 10,000 K.}\\
\multicolumn{4}{l}{\textbf{F126N, $\lambda_o$ = 1258.5 nm, $\Delta \lambda$ = 15.2 nm}} \\
\hline
\ion{Fe}{2} a$^4$D$_{7/2}$-a$^6$D$_{9/2}$ &1.257 $\mu$m & 1 & 1\\
\ion{Fe}{2} a$^4$D$_{1/2}$-a$^6$D$_{3/2}$ &1.252 $\mu$m & 0.003 & 0.01\\
\hline
\multicolumn{2}{l}{\textbf{F130N, $\lambda_o$ = 1300.0 nm, $\Delta \lambda$ = 15.6 nm}} \\
\hline
\ion{Fe}{2} a$^4$D$_{5/2}$-a$^6$D$_{5/2}$& 1.2946 $\mu$m & 0.05 & 0.17\\
\ion{Fe}{2} a$^4$D$_{3/2}$-a$^6$D$_{1/2}$& 1.2981 $\mu$m & 0.01 & 0.04\\
\hline
\multicolumn{2}{l}{\textbf{F164N, $\lambda_o$ = 1640.4 nm, $\Delta \lambda$ = 20.9 nm}} & 0.73 & 0.73\\
\hline
\ion{Fe}{2} a$^4$D$_{7/2}$-a$^4$F$_{9/2}$ & 1.6440 $\mu$m \\
\hline
\multicolumn{2}{l}{\textbf{F167N, $\lambda_o$ = 1664.2 nm, $\Delta \lambda$ = 21.0 nm}} \\
\hline
\ion{Fe}{2} a$^4$D$_{1/2}$-a$^4$F$_{5/2}$ &1.6642 $\mu$m & 0.013 & 0.05\\
\ion{Fe}{2} a$^4$D$_{5/2}$-a$^4$F$_{7/2}$ & 1.6773 $\mu$m & 0.04 & 0.1 \\
\hline
\end{tabular}
\end{table*}

\end{document}